\documentclass[aip, pop, reprint,numerical,letterpaper]{revtex4-1}

\usepackage{amsmath}				
\usepackage{amssymb}				
\usepackage{mathrsfs}				
\usepackage[mathscr]{euscript}		
\usepackage{amsthm}					
\usepackage{xspace}					
\usepackage{graphicx}
\usepackage[caption = false]{subfig}	
\usepackage[utf8]{inputenc}			
\usepackage{enumitem}				
\usepackage{color} 					
\usepackage{verbatim}				
\usepackage{tensor}					
\usepackage{easybmat}				
\usepackage{slashed}				

\usepackage{multirow}				
\usepackage{longtable}				
\usepackage{booktabs}				
\usepackage{tabularx}
\usepackage{siunitx}
\usepackage{csquotes}

\newcommand{\eq}[1]{(\ref{#1})}
\newcommand{\Eq}[1]{Eq.~(\ref{#1})}
\newcommand{\Eqs}[1]{Eqs.~(\ref{#1})}
\newcommand{\Fig}[1]{Fig.~\ref{#1}}

\newcommand{\Sec}[1]{Sec.~\ref{#1}}
\newcommand{\Tab}[1]{Table~\ref{#1}}
\newcommand{\Refa}[1]{Ref.~\onlinecite{#1}}
\newcommand{\Refs}[1]{Refs.~\onlinecite{#1}}
\newcommand{\App}[1]{Appendix~\ref{#1}}

\newcommand{\eg}{{e.g.,\/}\xspace}
\newcommand{\ie}{{i.e.,\/}\xspace}
\newcommand{\etal}{{\it et~al.\/}\xspace}
\newcommand{\pd}{\partial}

\newcommand{\dm}{\mathrm{d}}

\newcommand{\mhat}{\widehat{m}}
\newcommand{\I}{I_{\rm max}}

\newcommand{\Pref}{P_{\rm ref}}
\newcommand{\rhoref}{\rho_{\rm ref}}
\newcommand{\alpharef}{\alpha_{\rm ref}}
\newcommand{\Uref}{V_i}

\newcommand{\Dsb}{D_{\rm sb}}
\newcommand{\Psb}{P_{\rm sb}}
\newcommand{\Deltasb}{\Delta_{\rm sb}}
\newcommand{\Asb}{A_{\rm sb}}
\newcommand{\Msb}{M_{\rm sb}}

\newcommand{\const}{{\rm const.}}

\newcommand{\Rstar}{R_{\star}}

\newcommand{\Astar}{A_{\star}}
\newcommand{\Mstar}{M_{\star}}
\newcommand{\phistar}{\phi_{\star}}

\newcommand{\Rvc}{R_{\rm vc}}

\newcommand{\Dvc}{D_{\rm vc}}
\newcommand{\Pvc}{P_{\rm vc}}
\newcommand{\Deltavc}{\Delta_{\rm vc}}
\newcommand{\Avc}{A_{\rm vc}}

\newcommand{\Rstag}{R_{\rm stag}}

\newcommand{\Dstag}{D_{\rm stag}}
\newcommand{\Pstag}{P_{\rm stag}}
\newcommand{\Tstag}{T_{\rm stag}}

\newcommand{\mc}[1]{\mathcal{#1}}

\usepackage{hyperref}
\hypersetup{
  colorlinks   	= true, 	
  urlcolor     	= blue, 	
  linkcolor    	= blue, 	
  citecolor   	= blue 	
}
\begin{document}

\title{Asymptotic scaling laws for the stagnation conditions of Z-pinch implosions}

\author{D.~E.~Ruiz}
\email{deruiz@sandia.gov}
\author{C.~A.~Williams}
\author{R.~A.~Vesey}
\affiliation{Sandia National Laboratories, P.O. Box 5800, Albuquerque, New Mexico 87185-1186, USA}

\date{\today}


\begin{abstract}

Implosions of magnetically-driven annular shells (Z pinches) are studied in the laboratory to produce high-energy-density plasmas.  Such plasmas have a wide-range of applications including x-ray generation, controlled thermonuclear fusion, and astrophysics studies.  In this work, we theoretically investigate the in-flight dynamics of a magnetically-driven, imploding cylindrical shell that stagnates onto itself upon collision on axis.  The converging flow of the Z-pinch is analyzed by considering the implosion trajectory in the $(A, M)$ parametric plane, where $A$ is the in-flight aspect ratio and $M$ is the implosion Mach number.  For an ideal implosion in the absence of instabilities and in the limit of $A\gg1$, we derive asymptotic scaling laws for hydrodynamic quantities evaluated at stagnation (\eg density, temperature, and pressure) and for performance metrics (\eg soft x-ray emission, K-shell x-ray emission, and neutron yield) as functions of target-design parameters.

\end{abstract}

\maketitle

\section{Introduction}
\label{sec:intro}

For many decades, pulsed-power accelerators have been used for compressing electrical energy in space and time to implode cylindrical annular shells, which are often referred as Z pinches.\cite{Matzen:1999aa,ryutov2000,Matzen:2005aa,haines2011,mcbride2018,sinars2020}  Such magnetically-driven Z-pinch implosions are studied in the laboratory to produce high-energy-density (HED) plasmas.  Such plasmas have a wide-range of applications including x-ray generation,\cite{pereira1988,cuneo2005,schwarz2022,giuliani2015} controlled thermonuclear fusion,\cite{coverdale2007,Velikovich:2007hq,giuliani2015,Slutz:2012gp,Gomez:2014eta,Gomez:2020cd} and astrophysics studies.\cite{bailey2015,nagayama2019,valenzuela-villaseca2023}  Due to the aforementioned applications, it is important to understand how the hydrodynamic conditions assembled at stagnation depend on target-design parameters of a Z pinch.  

With this goal in mind, we study the idealized problem of a one-dimensional Z-pinch implosion in the absence of hydrodynamic instabilities and finite-conductivity effects.\cite{foot:diffusion}  More specifically, we investigate the in-flight dynamics of a magnetically-driven cylindrical shell, which then stagnates onto itself upon collision on axis.  The converging flow of the Z-pinch is analyzed by considering the implosion trajectory in the $(A, M)$ parametric plane, where $A$ is the in-flight aspect ratio (IFAR) and $M$ is the implosion Mach number.    The high--aspect-ratio limit $(A\gg1)$ is considered for most of the implosion trajectory.  We then derive asymptotic scaling laws for hydrodynamic conditions evaluated at stagnation (\eg density, temperature, and pressure) and performance metrics (\eg x-ray emission and neutron yield) as functions of target-design parameters, specifically the implosion velocity, the liner mass per-unit-length, and the liner entropy parameter.

Understanding the pressure amplification due to converging flows has been studied before in the HED literature.  In particular, this work follows the analysis presented in \Refa{basko2002} for imploding spherical shells driven by a constant pressure source, an idealized approximation often used for modeling laser-driven HED experiments. However, contrary to laser-driven HED experiments, the key feature of Z-pinch implosions is that the magnetic-drive pressure increases as the pinch radius decreases.  This distinction introduces new subtleties to the analysis of the shell in-flight dynamics and to the estimates of the plasma conditions assembled at stagnation.

The present work is organized as follows.  In \Sec{sec:kinematics}, we describe the general implosion kinematics of a Z pinch.  In \Sec{sec:inflight}, we discuss the in-flight shell dynamics of an imploding Z pinch.  More specifically, we analyze how the pressure, density, thickness, and aspect ratio of the cylindrical shell evolves as it converges on axis.  In \Sec{sec:scaling}, we derive scaling laws for the hydrodynamic quantities (\eg density, temperature, and pressure) evaluated at stagnation.  In \Sec{sec:perf}, we utilize the previous results to calculate the asymptotic dependencies of commonly referred performance metrics for Z-pinch applications (\eg x-ray emission and neutron yield) with respect to target-design parameters.  In \Sec{sec:discussion}, we discuss the results obtained.  In particular, we present how the hydrodynamic quantities at stagnation and the performance metrics scale when similarity scaling Z-pinch implosions to higher currents.  In \Sec{sec:conclusions}, we summarize our results.

\section{Implosion kinematics}
\label{sec:kinematics}

We consider the temporal evolution of a magnetically-driven cylindrical liner. To describe the kinematics of a Z-pinch implosion, we model the liner as a thin shell.\cite{rosenbluth1954,leontovich1956,ryutov2000,oreshkin2013}  Under the action of an external magnetic pressure, the governing equation for the radial motion of the liner is given by 
\begin{equation}
	\mhat \frac{\dm^2 R}{\dm t^2} = - 2 \pi R P_m (t),
	\label{eq:kinematics:R}
\end{equation}
where $R=R(t)$ is the outer radius of the liner, $\mhat$ is the liner mass per-unit-length, and $P_m(t) \doteq p_m\boldsymbol{(}t,R(t)\boldsymbol{)}$ is the magnetic pressure evaluated at the outer surface of the liner.  The liner mass per-unit-length $\mhat$ is defined as
\begin{equation}
	\mhat \doteq \pi \rho_0 \left( R_{\rm out,0}^2 - R_{\rm in,0}^2 \right),
	\label{eq:kinematics:mhat}
\end{equation}
where $\rho_0$ is the initial mass density of the liner, $R_{\rm out,0}$ is the initial outer radius, and $R_{\rm in,0}$ is the initial inner radius.  We denote $\Delta_0\doteq R_{\rm out,0}-R_{\rm in,0}$ as the initial thickness of the liner, so the initial aspect ratio is $A_0 \doteq R_{\rm out,0}/\Delta_0$.  For high--aspect-ratio liners, $A_0 \gg 1$.  Thus, the mass per-unit-length can be approximated by $\mhat \simeq 2 \pi \rho_0 R_0^2/A_0$, where we identified $R_0 \doteq R_{\rm out,0}$ for notational convenience.  

In \Eq{eq:kinematics:R}, the external magnetic pressure is given by
\begin{equation}
	p_m(t,r) \doteq \frac{B_\theta^2(t,r)}{2\mu_0} = \frac{\mu_0 I^2(t)}{8 \pi^2 r^2},
	\label{eq:kinematics:pm}
\end{equation}
where $B_\theta(t,r)$ is the azimuthal magnetic field, $I(t)$ is the electrical current passing through the shell, and $\mu_0$ is the magnetic permeability in free space.  In this work, we consider the following model for the current source $I(t)$.  Prior to $t=0$, we consider that the current has an initial ``foot" stage that is large enough in amplitude to launch a shock through the liner material (see \Sec{sec:phase1} for more details) but small enough such that the delivered kinetic energy to the liner is negligible.  After the foot stage, the current rises to its peak value $\I$ at $t=0$ and remains constant thereafter.  Up to this point, the liner motion is considered negligible so, from the perspective of the implosion kinematics, the current pulse is simply $I(t) = \I$ for $t\geq 0$.\cite{potter1978,foot:rise}

\begin{figure}
	\includegraphics[width=1\linewidth]{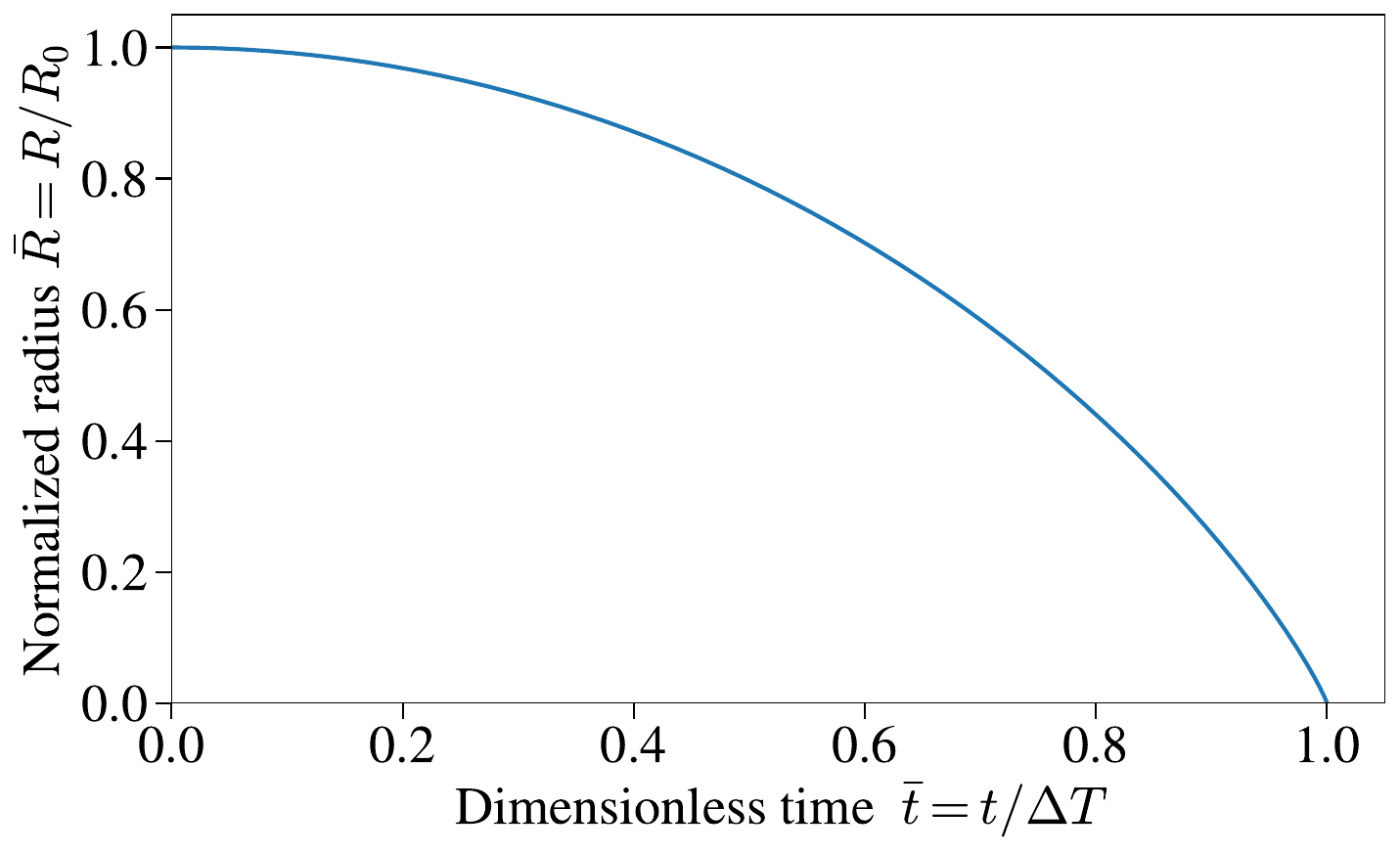}
	\caption{Implosion trajectory of a magnetically-driven cylindrical liner driven at constant current that is initially at rest.  The curve corresponds to the analytical solution in \Eq{eq:kinematics:Rsol}, and the implosion time $\Delta T$ is given by \Eq{eq:kinematics:DeltaT}.}
	\label{fig:R-T}
\end{figure}

\begin{figure}
	\includegraphics[width=1\linewidth]{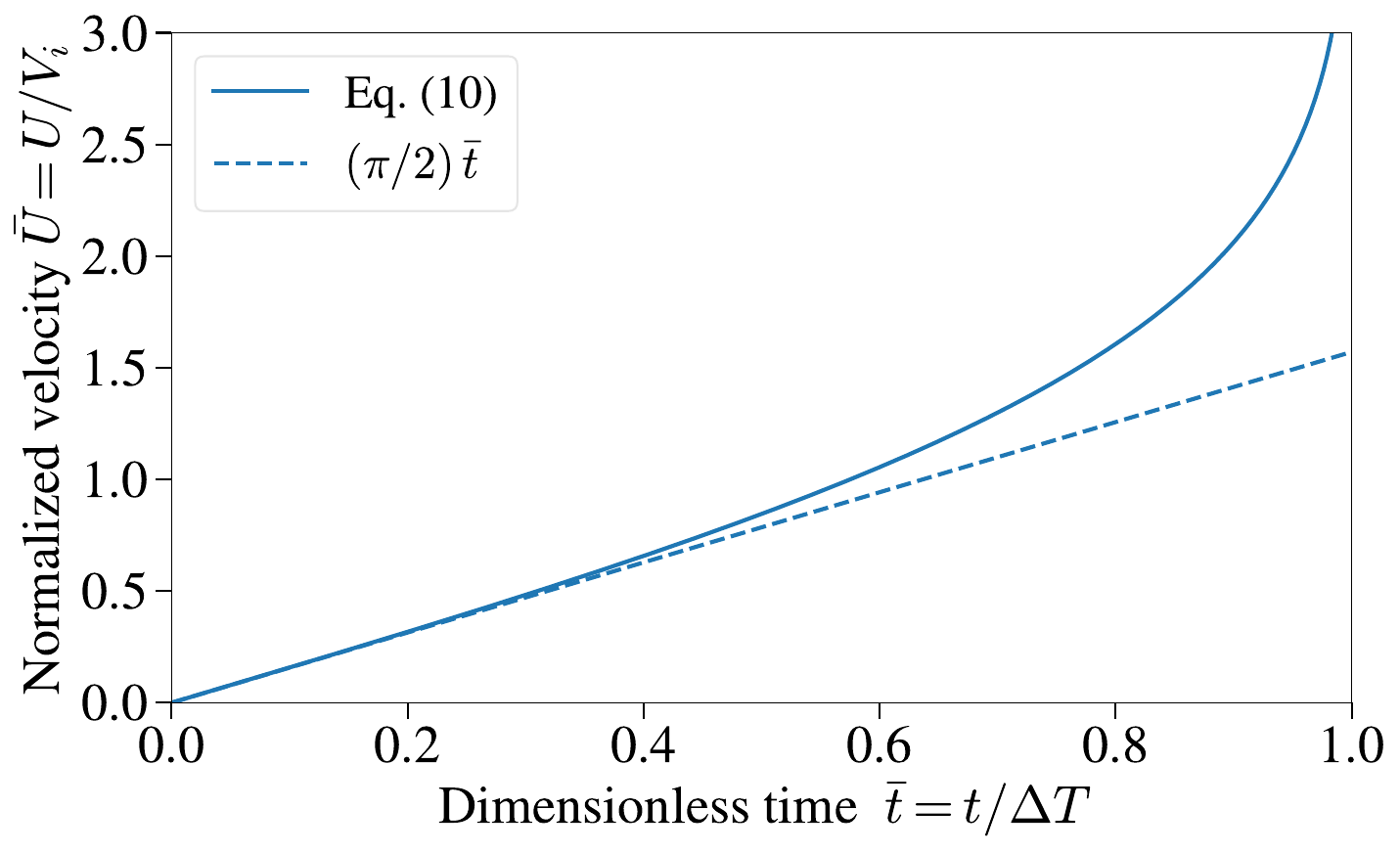}
	\caption{Dimensionless velocity of a magnetically-driven cylindrical liner driven at constant current that is initially at rest.}
	\label{fig:inverf}
\end{figure}

The implosion trajectory of the liner can be obtained by energy-conservation arguments.\cite{hussey1980,liberman1999}  When assuming that the velocity of the liner is negligible at $t=0$, we integrate \Eq{eq:kinematics:R} and obtain
\begin{equation}
	\left(\frac{\dm R}{\dm t} \right)^2 
			= \pi \left( \frac{R_0}{\Delta T} \right)^2 \ln \left( \frac{R_0}{R} \right),
	\label{eq:kinematics:U}
\end{equation}
where
\begin{equation}
	\Delta T 
		\doteq 	\left( \frac{\mhat}{4 P_m(0)} 	\right)^{1/2}
		=		\left( 2 \pi^2 \frac{\mhat R_0^2}{\mu_0 \I^2} \right)^{1/2}.
	\label{eq:kinematics:DeltaT}
\end{equation}

When taking the negative root of \Eq{eq:kinematics:U} and integrating once more, we obtain
\begin{equation}
	\frac{t}{\Delta T} 
			= - \frac{1}{\sqrt{\pi}}
				\int_1^{\bar{R}} \frac{\dm \bar{R}}{\sqrt{-\ln \bar{R}}},
	\label{eq:kinematics:Integral}
\end{equation}
where $\bar{R}(t) \doteq R(t)/R_0$ is the normalized pinch radius.  The integral on the right-hand side of \Eq{eq:kinematics:Integral} can be written in terms of the error function:
\begin{equation}
	\mathrm{erf}(x) = \frac{2}{\sqrt{\pi}} \int_0^x e^{-x^2} \, \dm x.
	\label{eq:kinematics:erf}
\end{equation}
This leads to
\begin{equation}
	\frac{t}{\Delta T} = \mathrm{erf}\left( \sqrt{- \ln \bar{R}} \right).
\end{equation}
We then introduce the inverse function erf$^{-1}(x)$ of the error function erf$(x)$ such that $\mathrm{erf}^{-1} \boldsymbol{(} \mathrm{erf}(x)\boldsymbol{)} = x$. The domain of erf$^{-1}(x)$ spans $x\in(-1,1)$ in the real line.  We then obtain a closed expression for the implosion trajectory of the Z-pinch:\cite{hussey1980}
\begin{equation}
	R(t) = R_0 \exp \left\{ - \left[\mathrm{erf}^{-1} \left( \frac{t}{\Delta T} \right) \right]^2 \right\}.
	\label{eq:kinematics:Rsol}
\end{equation}
Since erf$^{-1}(x)$ tends to infinity when its argument approaches unity, we conclude that $\Delta T$ in \Eq{eq:kinematics:DeltaT} corresponds to the implosion time of a Z pinch driven at constant current with zero initial velocity.  As shown in \Eq{eq:kinematics:DeltaT}, the implosion time increases for larger liner mass per-unit-length $\smash{\widehat{m}}$, larger initial radius $R_0$, and smaller peak current $\I$, as expected.  For illustration purposes, the radial implosion trajectory described by \Eq{eq:kinematics:Rsol} is shown in \Fig{fig:R-T}.

For the sake of completeness, we note that the obtained analytical solution in \Eq{eq:kinematics:Rsol} can be inserted into \Eq{eq:kinematics:U} to obtain the implosion velocity as a function of time:
\begin{equation}
	U(t) \doteq - \frac{\dm R}{\dm t}
			=	\sqrt{\pi} \frac{R_0}{\Delta T}
					\, \mathrm{erf}^{-1} \left( \frac{t}{\Delta T} \right) .
	\label{eq:kinematics:U2}
\end{equation}
The implosion velocity is plotted in \Fig{fig:inverf}.  For small times $t\ll \Delta T$, $U$ increases linearly in time; more specifically, $U\simeq (\pi/2)(R_0/\Delta T)(t/\Delta T)$.  According to \Eq{eq:kinematics:U2}, the implosion velocity theoretically diverges to infinity as $t/\Delta T \to 1$.  In reality, this does not occur due to the finite thickness of the shell, which limits the minimum radius to which current is delivered.  In addition, there can be current redistribution to larger radii caused by low-density plasmas,\cite{stollberg2023} an effect not considered in this study.

From \Eq{eq:kinematics:U2}, we identify a characteristic velocity
\begin{equation}
	\Uref \doteq \frac{R_0}{\Delta T} = \left( \frac{\mu_0 \I^2}{2 \pi^2 \mhat} \right)^{1/2}.
	\label{eq:kinematics:Vi}
\end{equation}
As expected, the characteristic velocity increases linearly with the peak current and is inversely proportional to the square root of the liner mass per-unit-length.  At fixed $\I$ and $R_0$, we note that the implosion velocity scales with the square root of the liner initial aspect ratio.

\section{Shell in-flight dynamics}
\label{sec:inflight}

For Z-pinch implosions, the dynamics of the in-flight shell conditions (notably the shell pressure, density, thickness, and aspect ratio) can be decomposed into four phases.  These phases are illustrated in \Fig{fig:A-M} and are denoted as the (i)~the magnetically-driven shock phase, (ii)~the isentropic-acceleration phase, (iii)~the acceleration-at-constant-shell-thickness phase, and (iv)~the void-closure phase. We now discuss how the in-flight shell conditions evolve during the implosion.

\subsection{Phase 1: Magnetically-driven shock}
\label{sec:phase1}

When a metallic liner is magnetically driven with a rapid rise in current and its thickness is sufficient for the characteristic sound waves to overlap each other within the liner, a shock wave is generated that propagates through the liner material.  In metals with finite electrical conductivity, this shock wave is followed by a nonlinear magnetic-field diffusion wave, which travels at anomalously high speeds due to the increase in metal resistivity with temperature. This paper will not delve into the complex dynamics occurring during this phase; for further details, we direct the reader to \Refs{knoepfel2000,krivosheev2011,oreshkin2012a,chaikovsky2015,oreshkin2024}.

However, one of the primary outcomes during this phase is the increase in entropy of the liner resulting from the passage of the shock wave.  When the shock has broken out of the inner surface of the liner, we assume that the liner material is in an isentropic state so that the fluid density $\rho(t,r)$ and pressure $p(t,r)$ within the interior of the liner are related by $p = \Pref ( \rho/\rhoref)^\gamma$.  Here $\Pref$ is a reference pressure, $\rhoref$ is a reference mass density, and $\gamma$ is an effective polytropic index.  More concretely, $\Pref$, $\rhoref$, and $\gamma$ are equation-of-state (EOS) parameters defining the post-shock isentrope of the material conditions inside the liner.  We define the entropy parameter $\alpharef \doteq \Pref / \rhoref^\gamma$, which we assume remains constant during the implosion until stagnation.  

It is worth noting that, when using an adiabatic EOS for the liner material, we are inherently ignoring energy-loss mechanisms, such as radiative losses, happening in-flight as the implosion proceeds.  This effect can be important for many radiating Z-pinch experiments.\cite{narkis2022}

In a 1D model, the fluid momentum equation within the interior of the liner can be written as
\begin{equation}
	\rho \left( \frac{\pd v_r}{\pd t} + v_r \frac{\pd}{\pd r} v_r \right) = - \frac{\pd p}{\pd r},
\end{equation}
where $v_r(t,r)$ is the radial velocity.  In the high--aspect-ratio limit, we may assume a uniform acceleration of all fluid elements within the liner.\cite{Nora:2014iq}  Therefore, the fluid velocity can be approximated by $v_r \simeq \dm R/ \dm t$, and
\begin{equation}
	\rho \frac{\dm^2 R}{\dm t^2} = - \frac{\pd p}{\pd r}.
	\label{eq:phase1:momentum}
\end{equation}
When substituting $p = \Pref ( \rho/\rhoref)^\gamma$ and integrating, we find the pressure field within the liner:\cite{Schmit:2020jd,ruiz2023}
\begin{equation}
	p(t,r) = P_m(t) \left( 1 + \frac{r-R}{\Delta} \right)^{\gamma/(\gamma-1)},
	\label{eq:phase1:pressure}
\end{equation}
where the liner in-flight thickness is given by
\begin{equation}
	\Delta(t) = \Delta_0 \frac{\gamma}{\gamma-1}
				\frac{\rho_0 R_0 }{\rhoref R(t)} 
				\left( \frac{\Pref}{P_m(t)}\right)^{1/\gamma}.
	\label{eq:phase1:Delta}
\end{equation}
Here we considered the region $r<R-\Delta$ as void with null pressure.  Also, $R$ corresponds to the outer liner surface, and the pressure there is equal to the magnetic pressure $P_m(t)$.  When using the EOS $p = \Pref ( \rho/\rhoref)^\gamma$, we obtain the density profile across the liner:
\begin{equation}
	\rho(t,r) = \rhoref \left( \frac{P_m}{\Pref}\right)^{1/\gamma} 
				\left( 1 + \frac{r-R}{\Delta} \right)^{1/(\gamma-1)}.
	\label{eq:phase1:rho}
\end{equation}

\begin{figure}
	\includegraphics[width=1\linewidth]{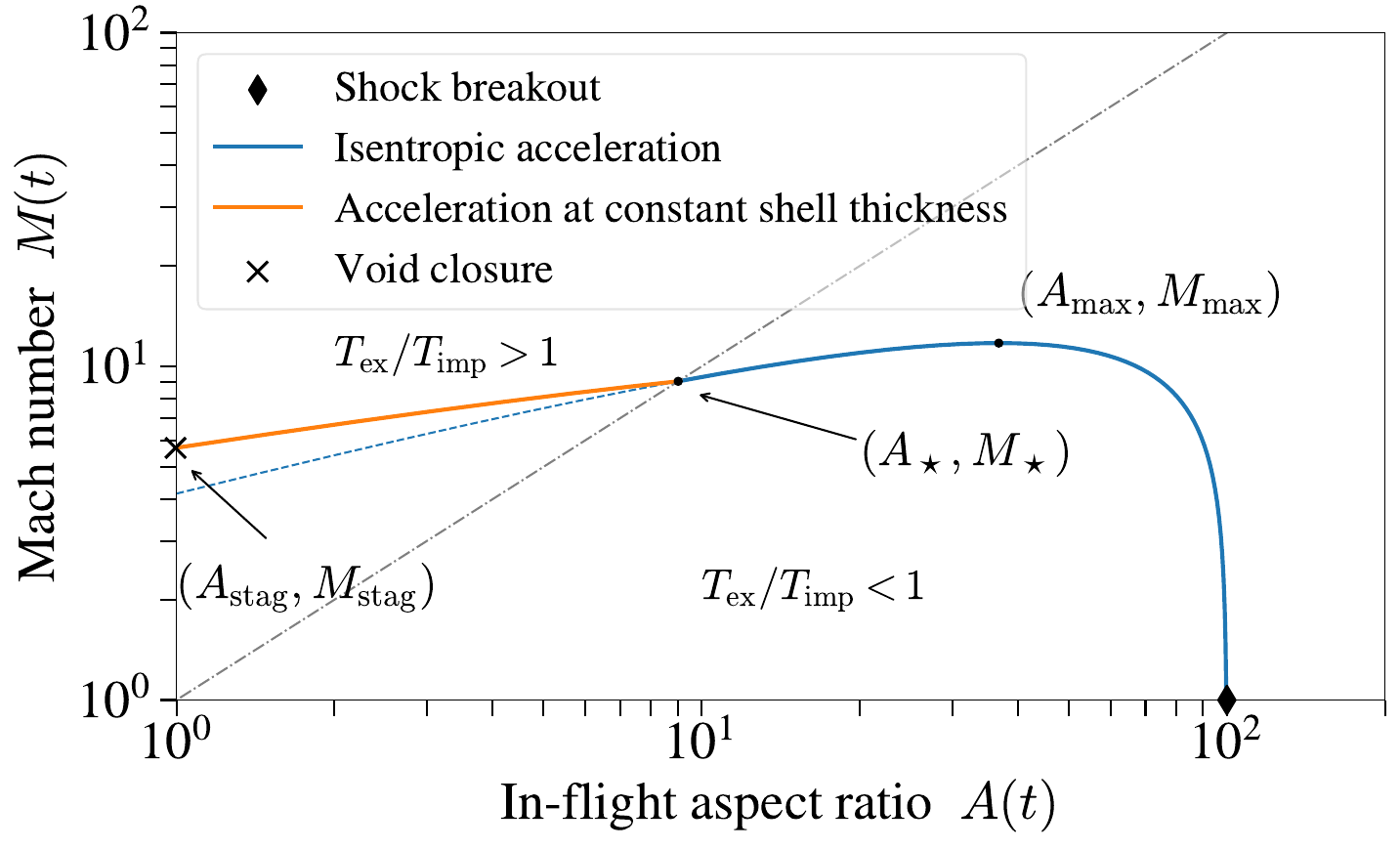}
	\caption{Example of an implosion trajectory of a magnetically-driven cylindrical liner in the $(A,M)$ parametric plane.  In this example, we consider $\Asb=10^2$ and $\gamma=5/3$.  The shock breakout point is plotted at $\Msb=1$.  We emphasize that the mathematical analysis presented in this work is an asymptotic theory valid in the high--aspect-ratio limit when $A(t) \gg 1$.  It is worth mentioning that high--aspect-ratio implosions of aluminum foils with $A_0\sim\mc{O}(10^3-10^4)$ have been experimentally and numerically studied at MA-scale pulsed-power machines.\cite{yager-elorriaga2015,yager-elorriaga2016,yager-elorriaga2018,sorokin2019,campbell2020,woolstrum2020,sorokin2022}}
	\label{fig:A-M}
\end{figure}

Let us now parameterize the dynamics of the imploding cylindrical shell in terms of five dynamical variables: the liner outer radius $R(t)$, the liner velocity $U(t)\doteq - \dot{R}$, the liner pressure $P(t)\doteq p(t,R)=P_m(t)$ evaluated at the outer surface, the liner density $D(t) \doteq \rho(t,R)$ at the outer surface, and the liner in-flight thickness $\Delta(t)$.  Following \Refa{basko2002}, we can construct two dimensionless parameters from the five variables $(R,U,P,D,\Delta)$.  These are the liner in-flight aspect ratio $A(t)$ and the liner Mach number $M(t)$, which are defined as
\begin{equation}
	A(t) \doteq \frac{R(t)}{\Delta(t)}, \qquad
	M(t) \doteq \frac{U(t)}{C(t)}.
	\label{eq:phase1:A-M}
\end{equation}
Here $C(t)$ is the characteristic sound speed within the liner and is given by
\begin{equation}
	C(t) \doteq \sqrt{\gamma P(t)/D(t)\,}.
	\label{eq:phase1:Cs}
\end{equation}
Based on these two dimensionless quantities, the liner in-flight dynamics can be described by trajectories in the $(A,M)$ parametric plane,\cite{basko2002} as shown in \Fig{fig:A-M}. 

To finalize this section, we consider that the liner IFAR $\Asb$ at shock breakout is known, where the ``sb" subscript denotes quantities evaluated at shock breakout.  Since the kinetic energy of the liner is negligible, we have $\Msb\simeq0$.  Upon using \Eqs{eq:kinematics:mhat}, \eq{eq:kinematics:pm}, and \eq{eq:phase1:Delta}, we find the following relationships:\cite{foot:simeq}
\begin{subequations}\label{eq:phase1:Qsb}
	\begin{align}
	\Psb &	\simeq 	\frac{\mu_0 \I^2}{8 \pi^2 R_0^2} 
				= 	\frac{\mhat}{4 \Delta T^2} , 
		\label{eq:phase1:Psb} \\
	\Dsb &= \rhoref \left( \frac{\Psb}{\Pref} \right)^{1/\gamma} = \frac{\Psb^{1/\gamma}}{\alpharef^{1/\gamma}},
		\label{eq:phase1:Dsb} \\
	\Deltasb &= \Delta_0 \frac{\gamma}{\gamma-1} \frac{\rho_0}{\Dsb}, 
		\label{eq:phase1:Deltasb}\\
	\Asb &\simeq \frac{R_0}{\Delta_{\rm sb}} = A_0 \frac{\gamma-1}{\gamma} \frac{\Dsb}{\rho_0}.
		\label{eq:phase1:Asb}
	\end{align}
\end{subequations}
Our next task will be to obtain a constitutive relationship between $A$ and $M$ for the next phases of the Z-pinch implosion shown in \Fig{fig:A-M}.  We shall use \Eqs{eq:phase1:Qsb} as ``initial conditions" for the in-flight dynamics.

\subsection{Phase 2: Isentropic acceleration}
\label{sec:phase2}

The next two phases of the liner dynamics can be distinguished by considering the ratio of the following two timescales.  The expansion timescale
\begin{equation}
	T_{\rm ex}(t) \doteq \frac{\Delta(t)}{C(t)}
	\label{eq:phase2:Tex}
\end{equation}
measures the characteristic time for a sound wave to traverse the liner shell.  In contrast, the implosion time
\begin{equation}
	T_{\rm imp}(t) 	\doteq	\frac{R(t)}{U(t)}
	\label{eq:phase2:Timp}
\end{equation}
measures the characteristic time for the change in the pinch radius.  The ratio $T_{\rm ex}/T_{\rm imp}$ can be written in terms of the liner IFAR and Mach number as follows:\cite{basko2002}
\begin{equation}
	\frac{T_{\rm ex}}{T_{\rm imp}}
		= 	\frac{\Delta}{R} \frac{U}{C} = \frac{M}{A}.
	\label{eq:phase2:ratioT}
\end{equation}

We use the ratio $T_{\rm ex}/T_{\rm imp} = M/A$ to differentiate between Phase 2 and Phase 3 of the liner implosion.  When $T_{\rm ex}/T_{\rm imp}  \ll 1$, the liner is moving at a sufficient slow velocity allowing sound waves to easily traverse the width of the liner.  In this regime, the isentropic state established by the shock in Phase 1 can be maintained in flight.  Hence, we designate Phase 2 as ``Isentropic acceleration".  In contrast, when $T_{\rm ex}/T_{\rm imp}  \gg 1$, sound waves traveling within the liner do not have sufficient time to fully traverse the shell width.  As further elaborated in \Sec{sec:phase3}, the shell thickness remains approximately constant in this regime.\cite{basko2002} Therefore, we refer to Phase 3 as ``Acceleration at constant shell thickness."  We shall now turn our attention to the dynamics occurring in Phase 2.

Since $M\simeq0$ at $t=0$, $T_{\rm ex}/T_{\rm imp} \simeq 0$ so the isentropic state is maintained.  In this phase, the liner pressure, density, thickness, and the IFAR vary with the implosion radius $R$ as
\begin{subequations}\label{eq:phase2:Q}
	\begin{align}
	\frac{P}{\Psb}	&=	\left(\frac{R}{R_0}\right)^{-2},
		\label{eq:phase2:P}\\
	\frac{D}{\Dsb}	&=	\left(\frac{R}{R_0}\right)^{-2/\gamma},
		\label{eq:phase2:Den}\\
	\frac{\Delta}{\Deltasb}	&=	\left(\frac{R}{R_0}\right)^{-1+2/\gamma},
		\label{eq:phase2:Delta} \\
	\frac{A}{\Asb}	&=	\left(\frac{R}{R_0}\right)^{2-2/\gamma}.
		\label{eq:phase2:A}
	\end{align}
\end{subequations}
As shown by \Eqs{eq:phase2:Q}, the liner pressure $P$ increases with smaller radius due to the increased magnetic pressure.  From \Eq{eq:phase2:Den}, the shell density increases to maintain the shell adiabat.  For $\gamma=5/3$, the liner thickness evolves as $\Delta\propto R^{1/5}$, a rather weak scaling with the liner radius.  This occurs due to the compensating effects of radial convergence (increases $\Delta$) and magnetic compression (reduces $\Delta$).  For $\gamma=5/3$, the IFAR follows $A\propto R^{4/5}$; \ie it decreases almost linearly with $R$.

Let us calculate the Mach number.  When substituting \Eqs{eq:kinematics:U}, \eq{eq:kinematics:DeltaT}, \eq{eq:phase1:Qsb}, and \eq{eq:phase2:Q} into \Eq{eq:phase1:A-M}, we obtain
\begin{align}
	M^2
		&	=	\frac{1}{\gamma \Psb/\Dsb}\frac{U^2}{(P/\Psb)/(D/\Dsb)} \notag \\
		&	=	\frac{\pi (R_0/\Delta T)^2}{\gamma \Psb/\Dsb}
				\left( 	\frac{R}{R_0} \right)^{2-2/\gamma}
		 		\ln \left( \frac{R_0}{R} \right) \notag \\
		&	=	\frac{2}{\gamma-1} \Asb
				\left( 	\frac{R}{R_0} \right)^{2-2/\gamma}
				\ln \left( \frac{R_0}{R} \right)  \notag \\
		&	=	\frac{\gamma}{(\gamma-1)^2}\Asb 
				\left( 	\frac{A}{\Asb} \right)
				\ln \left( \frac{\Asb}{A} \right),
	\label{eq:phase2:M2}
\end{align}
where we wrote the Mach number in terms of the IFAR using \Eq{eq:phase2:A} in the last line.

In \Fig{fig:A-M}, we plot the liner trajectory during this phase of the implosion.  As the IFAR decreases, the Mach number reaches a maximum.  From \Eq{eq:phase2:M2}, this occurs when $A=A_{\rm max} \doteq \Asb/e$, which is independent of the implosion trajectory or EOS properties.  As a result, the maximum Mach number is given by
\begin{equation}
	M_{\rm max} = \frac{\sqrt{\gamma \Asb}}{(\gamma-1)} \, e^{-1/2}.
	\label{eq:phase2:Mmax}
\end{equation}
Interestingly, the maximum liner Mach number scales with the square root of $\Asb$ as in spherical implosions.\cite{basko2002}  From \Eq{eq:phase2:A}, the radius at which the maximum Mach number occurs is given by
\begin{equation}
	R_{\rm max} \simeq R_0 \exp \left( -\frac{\gamma}{2\gamma-2} \right).
\end{equation}
For $\gamma=5/3$, we find $R_{\rm max}/R_1 \simeq 0.29$; \ie the liner converges by roughly a factor 3 to reach $M_{\rm max}$.

Regarding the ratio of the expansion time to the implosion time, we obtain the following expression:
\begin{equation}
	\frac{T_{\rm ex}}{T_{\rm imp}}
		=	\frac{1}{\gamma-1}\sqrt{\frac{\gamma}{\Asb}} 
				\left( 	\frac{\Asb}{A} \right)^{1/2}
				\ln^{1/2} \left( \frac{\Asb}{A} \right).
	\label{eq:phase2b:Tratio}
\end{equation}
As the liner radially implodes, the in-flight aspect ratio decreases, and the $T_{\rm ex}/T_{\rm imp}$ ratio increases.  Let $A_\star$ be the IFAR value at which the implosion time $T_{\rm imp}$ becomes comparable to the expansion time $T_{\rm ex}$, \ie when $T_{\rm ex}=T_{\rm imp}$.  Solving for $A_\star$ in \Eq{eq:phase2b:Tratio} leads to
\begin{equation}
	\frac{A_\star}{\Asb} 
			= \frac{1}{\phi_\star} W_0\left( \phi_\star \right).
	\label{eq:phase2:Astar}
\end{equation}
Here $W_0(x)$ is the principal branch of the Lambert $W$-function, which is defined as the solution of the transcendental equation $W(x)e^{W(x)}=x$.\cite{Hayes2005} We also introduced the parameter $\phi_\star$ given by
\begin{equation}
    \phistar \doteq \frac{(\gamma-1)^2}{\gamma} \Asb.
\end{equation}
In the high--aspect-ratio limit $(\Asb\gg1)$, $W_0(\phi_\star)$ has the following asymptotic dependency:
\begin{equation}
	W_0(\phi_\star) \simeq \ln \phi_\star - \ln \ln \phi_\star 
					 + \frac{\ln \ln \phi_\star}{\ln \phi_\star}.
	\label{eq:phase2:W0}
\end{equation}
In this limit, the dependency of $\Astar$ on $\Asb$ is only logarithmic to leading order.  As an example, when considering $\gamma=5/3$ and varying $\Asb$ between 10 and 1000 in \Eq{eq:phase2:Astar}, we obtain $\Astar\simeq 3.7$ and $\Astar\simeq 15.6$, respectively.

\begin{figure}
	\includegraphics[width=1\linewidth]{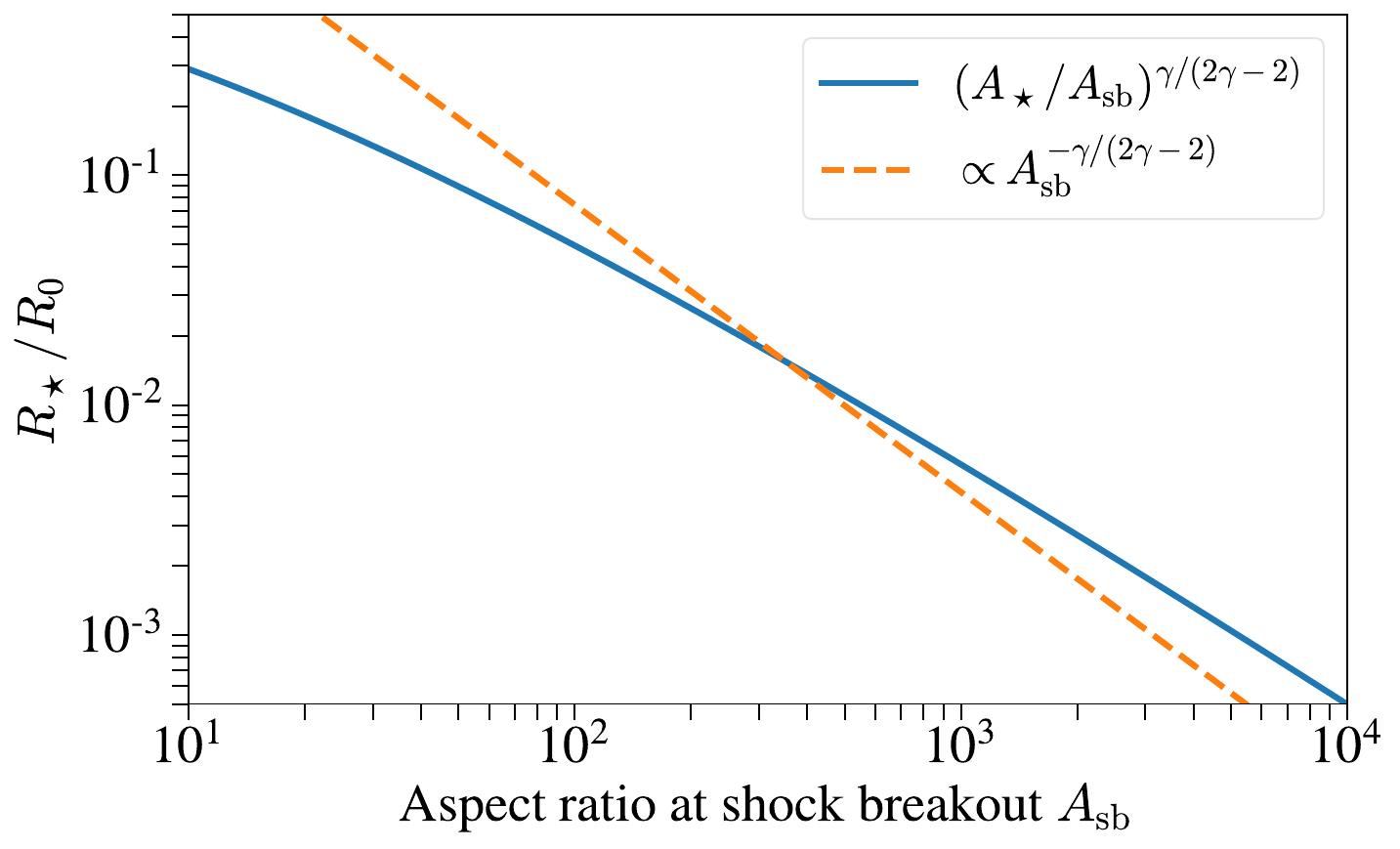}
	\caption{Normalized liner radius $R_\star/R_0$ as a function of $\Asb$ for the $\gamma=5/3$ case using \Eqs{eq:phase2:A} and \eq{eq:phase2:Astar}.  The solid blue line represents $\Rstar$ given in terms of the Lambert $W$-function $W_0$, and the dashed orange line represents the asymptotic expression of $\Rstar/R_0 \propto \Asb^{-\gamma/(2\gamma-2)}$ of the solution.}
	\label{fig:Rstar}
\end{figure}

To conclude this section, it is worth noting that the transition $T_{\rm ex}=T_{\rm imp}$ occurs relatively late in a Z-pinch implosion.  Upon substituting \Eq{eq:phase2:A} into \Eq{eq:phase2:Astar}, we can calculate the radius $\Rstar$.  As observed in \Fig{fig:Rstar}, $T_{\rm ex}=T_{\rm imp}$ happens almost near the point of stagnation since $\Rstar/R_0\ll1$.  For example, for $\Asb=100$ and $\gamma=5/3$, we obtain $\Rstar/R_0\simeq0.05$; \ie the transition occurs at a convergence ratio of approximately 20.  The occurrence of the transition $T_{\rm ex}=T_{\rm imp}$ at a relatively late stage during a Z-pinch implosion contrasts with spherical implosions, in which this transition takes place at an earlier stage.  This point will be further discussed in \Sec{sec:sphere}.

\subsection{Phase 3: Acceleration at constant shell thickness}
\label{sec:phase3}

In the next phase of the shell in-flight dynamics, $T_{\rm ex}\gg T_{\rm imp}$.  In this regime, the shell is moving at a considerable velocity not allowing sound waves to completely traverse the width of the shell.  As shown in \App{app:Delta}, the shell thickness $\Delta$ remains approximately constant during this phase of the implosion.\cite{basko2002}  Hence, the liner pressure, density, thickness, and the IFAR approximately obey
\begin{subequations}\label{eq:phase3:Q}
	\begin{align}
	\frac{P}{P_\star}	&=	\left(\frac{R}{R_\star}\right)^{-\gamma}, 
			\label{eq:phase3:P}	\\
	\frac{D}{D_\star}	&=	\left(\frac{R}{R_\star}\right)^{-1}, 
			\label{eq:phase3:Den}\\
	\Delta				&=	\Delta_\star ,
			\label{eq:phase3:Delta}\\
	\frac{A}{A_\star}	&=	\frac{R}{R_\star},
			\label{eq:phase3:A}
	\end{align}
\end{subequations}
where the ``$\star$" subscript denotes quantities evaluated at $T_{\rm ex}/T_{\rm imp}=1$ using \Eqs{eq:phase2:Q}.  Due to mass conservation, the density of the shell is inversely proportional to the shell radius meaning that the shell density rapidly increases during this phase of the implosion.  To maintain the in-flight adiabat, the pressure must behave as $P\propto D^\gamma$ leading to \Eq{eq:phase3:P}.  Because $\Delta=\const$, the liner IFAR is proportional to the liner radius.

Concerning the Mach number, we find
\begin{align}
	\frac{M^2}{M_\star^2}
		&	=	\frac{U^2/U_\star^2}{ \left(P/P_\star\right)^{1-1/\gamma}} 
			=	\frac{\ln \left( R_0/R \right)}{\ln \left( R_0/R_\star \right)}
		 		\left( 	\frac{R}{R_\star} \right)^{\gamma-1} 
		 		\notag \\
		&	=	\left[ 1- \frac{\ln \left(R/\Rstar \right)}{\ln \left( R_0/R_\star \right)} \right]
		 		\left( 	\frac{R}{R_\star} \right)^{\gamma-1} 
		 		\notag \\
		&	=	\left[ 1- \frac{\ln \left(A/\Astar \right)}{\ln \left( R_0/R_\star \right)} \right]
		 		\left( 	\frac{A}{A_\star} \right)^{\gamma-1} .
	\label{eq:phase3:M2}
\end{align}
An example trajectory of the Mach number as a function of the IFAR is shown in \Fig{fig:A-M} (orange line).  As the shell IFAR decreases, the Mach number also decreases.  However, the rate of decrease is smaller than that computed using \Eq{eq:phase2:M2}.  (For comparison, see the blue-dashed line in \Fig{fig:A-M}.)  

\begin{figure}
	\includegraphics[width=1\linewidth]{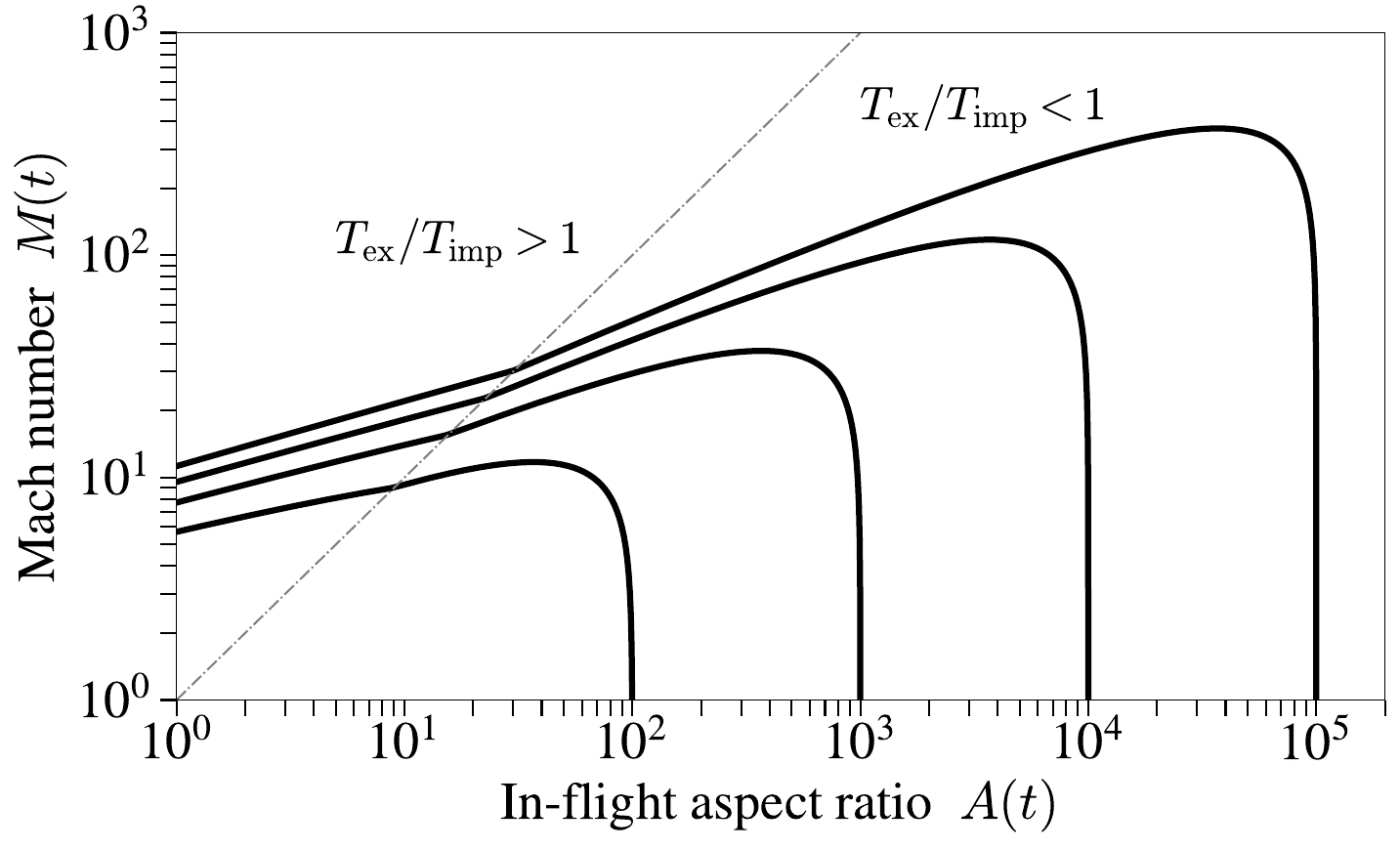}
	\caption{Implosion trajectories in the $(A,M)$ space for different $A_{\rm sb}$ values. We consider $\gamma=5/3$.}
	\label{fig:A-M2}
\end{figure}

\subsection{Phase 4: Void closure}
\label{sec:phase4}

The ``void closure" phase concerns the moment before the liner bounces off axis and the stagnation shock is launched within the liner material.  Here the aspect ratio of the liner is nearly unity so that $A\simeq1$.  Hence, the radius $\Rvc$ of the liner at void closure is well approximated by $\Rvc \simeq \Deltavc = \Delta_*$, where we used \Eq{eq:phase3:Delta}.  Since the hydrodynamic relations in \Eqs{eq:phase3:Q} depend on the ratio $R/R_\star$, we obtain $\Rvc /R_\star \simeq \Delta_\star/R_\star = 1/A_\star$ at the moment of void closure.  In other words, the shell hydrodynamic quantities at void closure only depend on the shell IFAR evaluated at the transition when $T_{\rm ex} = T_{\rm imp}$!  When using \Eqs{eq:phase3:Q}, we obtain
\begin{subequations}\label{eq:phase4:Q}
	\begin{align}
	\Pvc		&=	P_\star A_\star^\gamma, 
			\label{eq:phase4:P}	\\
	\Dvc		&=	D_\star A_\star, 
			\label{eq:phase4:Den}\\
	\Deltavc	&=	\Delta_\star ,
			\label{eq:phase4:Delta}\\
	\Avc		&=	1,
			\label{eq:phase4:A}
	\end{align}
\end{subequations}
where the subscript ``vc" denotes quantities evaluated at the moment of void closure.

Figure \ref{fig:A-M2} illustrates several implosion trajectories in the $(A,M)$ plane, where the shell aspect ratio $\Asb$ at shock breakout is varied. After the initial acceleration phase, the implosion Mach number $M$ is much greater than unity for the majority of the implosion duration.  As shown, the maximum Mach number increases as the square root of $\Asb$.  After the maximum Mach number has been achieved, the examples shown follow similar implosion trajectories.  Finally, $M(A=1)$ only varies by a few units even when changing the initial aspect ratio by several orders of magnitude.  A similar effect occurs in spherical implosions (cf. Figure 1 of \Refa{basko2002}).

\section{Scaling laws for stagnation conditions with target-design parameters}
\label{sec:scaling}

Once the shell has collided onto itself on axis, it is reasonable to assume that the internal energy of the plasma column will scale as the kinetic energy of the shell at the time of stagnation.  Therefore, we have
\begin{equation}
	\widehat{\mc{E}}_{\rm int} 
		\sim \mhat U_{\rm stag}^2 
		\sim \mhat \left( \frac{R_0}{\Delta T} \right)^2 \ln \left( \frac{R_0}{\Rstag} \right),
	\label{eq:scaling:Eint_aux}
\end{equation}
where $\widehat{\mc{E}}_{\rm int}$ is the internal energy per-unit-length of the plasma column.  Based on the discussion in \Sec{sec:phase4}, the stagnation radius $\Rstag$ follows $\Rstag\simeq \Delta_\star$.  The shell thickness at stagnation follows the scaling
\begin{equation}
	\Delta_\star
		=	R_0 \frac{\Rstar}{R_0} \frac{\Delta_\star}{\Rstar}
		=	R_0 \left(\frac{\Astar}{\Asb}\right)^{5/4} \frac{1}{\Astar}
		\sim \frac{R_0}{\Asb^{5/4}},
	\label{eq:scaling:Deltastar}
\end{equation}
where we used \Eq{eq:phase2:A}.  In the last relation, we used the results given in \Eqs{eq:phase2:Astar}--\eq{eq:phase2:W0}, which show that $\Astar$ is a weak function of the liner aspect ratio at shock breakout.  For the sake of simplicity, here and in the upcoming calculations, we shall focus on the $\gamma=5/3$ case.  Equation \eq{eq:scaling:Deltastar} shows that the stagnation radius scales proportionally to the initial radius of the Z pinch and is closely inverse proportional to the aspect ratio $\Asb$ at shock breakout.  When using \Eqs{eq:phase1:Qsb}, we note that $\Asb$ scales as
\begin{equation}
	\Asb \sim \frac{R_0^{4/5} V_i^{6/5}}{\mhat^{2/5} \alpharef^{3/5}},
	\label{eq:scaling:Asb}
\end{equation}
where $\Uref \doteq R_0/\Delta T$ is the characteristic implosion velocity defined in \Eq{eq:kinematics:Vi}.  When substituting \Eq{eq:scaling:Asb} into $\smash{\Rstag\simeq \Rvc \sim R_0 \Asb^{-5/4}}$, we find that the pinch radius $\Rstag$ at stagnation scales as
\begin{equation}
	\Rstag \sim \frac{\alpharef^{3/4} \mhat^{1/2}}{\Uref^{3/2}}.
	\label{eq:scaling:Rstag}
\end{equation}
Hence, to obtain more compact stagnation columns, it is necessary to increase the implosion velocity, lower the in-flight entropy parameter, and implode less massive shells.

When substituting \Eq{eq:scaling:Rstag} into \Eq{eq:scaling:Eint_aux}, we find that the scaling of the internal energy per-unit-length of the plasma column approximately obeys
\begin{align}
	\widehat{\mc{E}}_{\rm int} 
		&	\sim \mhat \Uref^2 \ln\left( \Asb^{5/4} \right)
		\sim \I^2 \ln\left( \frac{R_0 V_i^{3/2}}{\mhat^{1/2} \alpharef^{3/4}} \right),
	\label{eq:scaling:Eint}
\end{align}
where we substituted \Eq{eq:kinematics:Vi}.  Unsurprisingly, we find that $\smash{\widehat{\mc{E}}_{\rm int}}$ primarily scales with the peak current squared applied to the Z pinch.  Equation \ref{eq:scaling:Eint} also suggests that more energy can be delivered to Z pinches with higher aspect ratios.  However, the dependency on the aspect ratio $\Asb$ is only logarithmic.  Although it is important to be aware of the logarithmic dependency on $\Asb$, we shall ignore this contribution in the ensuing calculations.

It is worth mentioning that the exact constant of proportionality between the internal energy of the plasma column and the kinetic energy of the shell prior to void closure in \Eq{eq:scaling:Eint} is likely a complicated function of the shell adiabat and the spatial distribution of mass and energy within the shell. Furthermore, other non-ideal effects, such as the additional compression of the column by the magnetic field following void closure and energy leakage due to radiation losses in strongly radiating Z-pinches, will influence the scaling of the internal energy of the plasma column. Calculating the corrections associated with the aforementioned effects is beyond the scope of this paper.

To obtain the scaling of the plasma pressure at stagnation, we assume that the thermalization of the shell kinetic energy occurs at a radius $\Rstag$.  Therefore, 
\begin{equation}
	\Pstag
		\sim \frac{\widehat{\mc{E}}_{\rm int}}{\Rstag^2}
		\sim \frac{\Uref^5}{\alpharef^{3/2}} ,
	\label{eq:scaling:Pstag}
\end{equation}
where we used \Eq{eq:scaling:Rstag}.  From \Eq{eq:scaling:Pstag}, it follows that higher stagnation pressures can be achieved by mainly increasing the implosion velocity or by decreasing the entropy parameter $\alpharef$.  According to this estimate, the shell mass per-unit-length does not contribute to the scaling in pressure.

Based on mass conservation, the density of the plasma column at stagnation should obey $\Dstag \Rstag^2 \sim \mhat$.  Upon substitution of \Eq{eq:scaling:Rstag}, we obtain
\begin{equation}
	\Dstag
		\sim \frac{\mhat}{\Rstag^2}
		\sim \frac{\Uref^3}{\alpharef^{3/2}} .
	\label{eq:scaling:Dstag}
\end{equation}
Hence, the shell density scales as $\Uref^3$ with the implosion velocity and as $\smash{\alpharef^{-3/2}}$ with the shell entropy parameter.  As for the pressure at stagnation, the plasma density does not show any dependency with the liner mass per-unit-length (within the accuracy of this model).

Regarding the characteristic temperature of the shell near stagnation, we pose that the temperature follows the scaling corresponding to an ideal gas such that 
\begin{equation}
	\Tstag
		\sim \frac{\Pstag}{\Dstag}
		\sim \Uref^2 .
	\label{eq:scaling:Tstag}
\end{equation}
In other words, the temperature scales with the implosion velocity squared.  It is interesting to note that, within this model, the stagnation temperature has no dependency on the shell in-flight entropy parameter.

Other important metrics for Z-pinch x-ray and neutron sources are the areal density $\sigma_{\rm stag}$ and the confinement time $\tau$ at stagnation.  We expect that the areal density will obey the scaling
\begin{equation}
	\sigma_{\rm stag} \sim \Dstag \Rstag
			\sim \frac{\mhat^{1/2} \Uref^{3/2}}{\alpharef^{3/4} } 	.
	\label{eq:scaling:sigma}
\end{equation}
We propose that the confinement time scales as the characteristic radius of the stagnated plasma column divided by the characteristic implosion velocity.  This gives
\begin{equation}
	\tau_{\rm stag} 	\sim \frac{\Rstag}{\Uref}
			\sim \frac{\alpharef^{3/4} \mhat^{1/2}}{\Uref^{5/2}} .
	\label{eq:scaling:tau}
\end{equation}
The scaling laws derived in \Eqs{eq:scaling:Rstag}--\eq{eq:scaling:tau} suggest that the implosion velocity is the most important lever for enhancing the HED conditions of the Z-pinch plasma at stagnation.  We note that the $\Uref^5$ dependency of the plasma pressure in \Eq{eq:scaling:Pstag} represents a stronger scaling in velocity than that for spherical laser-driven implosions, where the scaling is $\Pstag \sim \Uref^4$ (see \Refa{basko2002}).  The next most important parameter is the shell entropy parameter $\alpharef$.  When lowered, it can increase the density, pressure, and areal density of the stagnated plasma column.  Finally, increasing the mass per-unit-length $\mhat$ is a useful design tool for increasing the pinch radius at stagnation, the areal density, and the confinement time.  

\section{Scaling laws for performance metrics}
\label{sec:perf}

\subsection{X-ray emission by Bremsstrahlung}
\label{sec:x-ray}

One application of Z pinches, such as wire arrays\cite{cuneo2005,schwarz2022} and gas puffs\cite{giuliani2015}, is the generation of x rays. Free-free electron bremsstrahlung emission is the dominant mechanism for soft x-ray emission for strongly ionized, low atomic number plasmas. If the stagnated plasma column is optically thin, the soft x-ray energy emission per-unit-length of a hot, uniform, cylindrical plasma is\cite{Atzeni:2009phys} 
\begin{equation}
	\widehat{\mc{P}}_{\rm ff}
		\doteq  \theta_{\rm ff}
					\sqrt{\frac{k_\text{B} \Tstag}{m_e c^2}} \frac{Z^3 \Dstag^2}{m_i^2}
					\pi R_{\rm stag}^2 \tau_{\rm stag},
	\label{eq:perf:xray}
\end{equation}
where
\begin{equation}
    \theta_{\rm ff} \doteq
        \frac{32}{3 \sqrt{6 \pi}}
        \bar{g} \frac{\alpha^3 \hbar^2 c}{m_e}.
\end{equation}
In the above, $m_e$ is the electron mass, $m_i$ is the mass of the ion species composing the Z-pinch liner, $Z$ is the average ion charge, $\bar{g}$ is the Gaunt factor (typically of order unity), $k_{\rm B}$ is Boltzmann's constant, $\hbar\doteq h/(2\pi)$ is the Planck constant, $c$ is the speed of light, and $\alpha \simeq 1/137$ is the fine-structure constant.  The factor $\pi R_{\rm stag}^2$ denotes the cross-sectional area of the plasma column.  When substituting the results from \Sec{sec:scaling} into \Eq{eq:perf:xray}, we find
%
%
%
\begin{equation}
	\widehat{\mc{P}}_{\rm ff}
		\sim \Dstag^2 \Tstag^{1/2} \Rstag^2 \tau_{\rm stag}
		\sim \frac{\mhat^{3/2} \Uref^{3/2} }{\alpharef^{3/4}}.
	\label{eq:perf:xray1}
\end{equation}

Based on \Eq{eq:kinematics:Vi}, the product $\mhat V_i^2$ is constant when keeping the peak current $\I$ fixed. In this case, the x-ray energy emission by Bremsstrahlung follows 
\begin{equation}
	\widehat{\mc{P}}_{\rm ff}
		\sim \left(\frac{\mhat}{\alpharef}\right)^{3/4}
	\label{eq:perf:xray2}
\end{equation}
for $\I=\const$, which shows that x-ray emission increases with the mass per-unit-length.  In this vein, M.~Gersten \etal reported experiments of imploding Al wire arrays in which the radius and mass per-unit-length were scaled to approximately satisfy $\mhat R_0^2 =\const$\cite{gersten1986}  In those experiments, the smaller-radius, larger-mass targets led to lower-temperature and higher-density stagnated plasma columns.  (In accordance to the scaling laws for the stagnation conditions presented in \Sec{sec:scaling}.)  In particular, the $\sim$5x more massive implosion produced $\sim$30x more x-ray yield (above 1 keV) than the control target with nominal mass per-unit-length.  When using \Eq{eq:perf:xray2}, we find an expected increase of only 3x.  It is reassuring that \Eq{eq:perf:xray2} qualitatively reproduces the observed trend in performance; however, the observed increase in performance in the experiments is much higher than what theory suggests.  Possible reasons for the discrepancy are that the larger-in-radius (less massive) targets are more prone to in-flight instabilities, have higher kinetic energy per nuclei at stagnation, and could be more susceptible to turbulence heating.\cite{pereira1988}  These non-ideal effects could result in larger, higher-temperature, lower-density plasma columns with lower emissivity at stagnation.

\subsection{X-ray emission by bound-bound transitions}
\label{sec:line emission}


%
In a fully ionized plasma, x rays are emitted in a continuous spectrum, reflecting the fact that the emitted photons originate from free electrons which begin in one plane-wave state and end in a different plane-wave state as a result of the emission. By contrast, a plasma in which the ionization is incomplete harbors atoms with electrons that remain in various bound states. Transitions between two of these discrete states results in emission lines characteristic of the radiating element. Many Z-pinch implosions are geared toward generating conditions under which copious amounts of He-$\alpha$ x rays are produced (that is, x rays engendered by transitions from the atomic L shell to the K shell in atoms ionized down to the final two electrons). This is the dominant form of soft x-ray emission in partially ionized mid-atomic number plasmas. The emitted energy per-unit-length resulting from bound-bound transitions from principal quantum number $m \rightarrow n$ in a cylindrical stagnation column is given by\cite{griem1997}

\begin{equation}
	\widehat{\mc{P}}_{\rm bb}
		=  \theta_{\rm bb} \sqrt{\frac{m_ec^2}{k_{\rm B} \Tstag}} 
            \exp\left(\frac{-E_{mn}}{k_{\rm B}\Tstag}\right) 
            \frac{Z \Dstag^2 }{m_i^2} \pi \Rstag^2 \tau_{\rm stag},
	\label{eq:perf:bound-bound form}
\end{equation}
where
\begin{equation}
	\theta_{\rm bb}
		\doteq 8 \pi \sqrt{\frac{\pi}{6}} 
        \bar{g} f_{mn}
        \frac{\alpha^2 \hbar^2 c}{m_e},
	\label{eq:perf:bound-bound_factors}
\end{equation}
$E_{mn}$ is the energy of the emitted photon, and $f_{mn}$ is the oscillator strength of the transition.

The temperature dependence of \Eq{eq:perf:bound-bound form} can be simplified by imposing a power law form and solving for the power index that satisfies the imposed identity. That is, we seek the $n$ that satisfies
\begin{equation}
    \frac{\widehat{\mc{P}}_{\rm bb}(T)}{\widehat{\mc{P}}_{\rm bb}(T_0)} \simeq \bigg( \frac{T}{T_0}\bigg)^n ,
    \label{eq:perf:power law}
\end{equation}
for temperatures in the vicinity of some desired $T_0$.

To isolate the temperature dependence of \Eq{eq:perf:bound-bound form}, we group the temperature-independent factors into a constant $b$ and define $\delta \doteq E_{mn}/k_\text{B}$ so that \Eq{eq:perf:bound-bound form} becomes
\begin{equation}
	\widehat{\mc{P}}_{\rm bb} (T)
		= \frac{b}{\sqrt{T}} e^{-\delta/T}.
	\label{eq:perf:bb working form}
\end{equation}
Solving for $n$ in \Eqs{eq:perf:power law} and \eq{eq:perf:bb working form} yields
\begin{equation}
    n   = \frac{\ln{\Big(\frac{\widehat{\mc{P}}_{\rm bb}(T)}{\widehat{\mc{P}}_{\rm bb}(T_0)}\Big)}}{\ln{\Big(\frac{T}{T_0}\Big)}} 
        = \frac{\frac{\delta}{T} [\frac{T}{T_0} -1]}{\ln{\Big(\frac{T}{T_0}\Big)}} - \frac{1}{2},
	\label{eq:perf:n as ratio}
\end{equation}
The first term of the Taylor expansion of $\ln{(T/T_0})$ is $(T/T_0 -1)$. Therefore, the power index itself is a function of temperature and equal (in first order) to
\begin{equation}
    n = \frac{\delta}{T} - \frac{1}{2}.
    \label{eq:perf:final n}
\end{equation}
Note that this scaling is only approximate in the sense that any calculation of the x-ray yield increase resulting from an increase in temperature necessarily spans multiple temperatures, all with different power indices. When calculating the scaling at a given temperature $T_0$, however, \Eq{eq:perf:final n} is exact. This is because $\ln{x}\simeq(x-1)$ when $x\simeq1$, so keeping only the first term in the Taylor expansion yields no error, and the slope of the \enquote{true} function \eq{eq:perf:bb working form} exactly matches that of $T^n$.  The region around $T_0$ over which this scaling is valid depends ultimately on how quickly the other terms in the expansion begin to contribute, which generally manifests in relatively small windows for $T \ll \delta$, and large windows when $T \simeq \delta$. The emission is roughly constant when the plasma temperature is comparable to the energy of the line of interest, as can be observed from \Eq{eq:perf:final n}, resulting in a large temperature range over which \Eq{eq:perf:final n} applies. For example, in a $\SI{1.85}{\kilo \electronvolt}$ neon plasma—whose He-$\alpha$ line is $\SI{0.922}{\kilo \electronvolt}$—the emission is constant with respect to temperature for $\SI{1.08}{\kilo \electronvolt} \leq T \leq \SI{4.35}{\kilo \electronvolt}$, while a krypton plasma (He-$\alpha$ line of $\SI{13.114}{\kilo \electronvolt}$) of equal temperature has $\smash{\widehat{\mc{P}}_{\rm bb} \propto T^{6.6}}$ for $\SI{1.57}{\kilo \electronvolt} \leq T \leq \SI{2.22}{\kilo \electronvolt}$, where the bounds indicate the region over which the approximate function is within $\SI{10}{\percent}$ of the exact value.

The preceding discussion makes clear that the scaling of bound-bound emission with temperature depends intimately on which lines are being pursued and which temperature ranges are accessible with the Z-pinch. In reality, a wide range of plasma parameters can be realized depending on the specific scheme being employed, so in order to ultimately find a proper current scaling we must apply constraints that do not pertain to all Z-pinches. As an exemplary plasma, we consider only the $\SI{3.14}{\kilo \electronvolt}$ He-$\alpha$ line of argon at $\Tstag = \SI{1.85}{\kilo \electronvolt}$,\cite{jones2014} yielding $\smash{\widehat{\mc{P}}_{\rm bb} \propto \Dstag^2 \Tstag^{1.2} \Rstag^2 \tau_{\rm stag}}$. We can use the results from \Sec{sec:scaling} to conclude
\begin{equation}
    \widehat{\mc{P}}_{\rm bb} \sim \frac{\mhat^{3/2}}{\alpharef^{3/4}} \Uref^{2.9}.
\end{equation}
Despite having a rather distinct origin from the bremsstrahlung x-ray production, x-ray line emission shares a similar scaling relationship with respect to the mass per-unit-length and the entropy parameter.  However, the dependency on the shell velocity is stronger for line emission as compared to bremsstrahlung emission.



\subsection{Neutron yield}
\label{sec:neutron}

Another studied application of Z pinches is the production of neutrons via nuclear fusion.  One widely studied Z-pinch neutron source is the gas puff.\cite{coverdale2007,Velikovich:2007hq}  For a Z pinch liner composed of equimolar DT fuel, the neutron yield per-unit-length is approximately given by\cite{Atzeni:2009phys}
\begin{equation}
	\widehat{Y}
		=	\frac{1}{4} \left( \frac{\Dstag}{m_i} \right)^2 \, \langle \sigma v \rangle_{\rm DT} \, \pi \Rstag^2  \tau_{\rm stag},
	\label{eq:perf:yield}
\end{equation}
where $\langle \sigma v \rangle_{\rm DT}$ is the DT fusion reactivity.  Between the 3- and 5-keV temperature range, the DT fusion reactivity can be approximated by $\langle \sigma v \rangle_{\rm DT} \sim T^4$.\cite{Bosch:1992aa,hurricane2016a}  When substituting the results from \Sec{sec:scaling} into \Eq{eq:perf:yield}, we find
\begin{equation}
	\widehat{Y}
		\sim \Dstag^2 \Tstag^{4} \Rstag^2 \tau
		\sim \frac{\mhat^{3/2} \Uref^{17/2} }{\alpharef^{3/4}}.
\end{equation} 
In this case, due to the strong dependency with the ion temperature at stagnation, the neutron yield scales strongly with the characteristic implosion velocity $\Uref$.

It is interesting to note that, for both x-ray emission and neutron yield, there is a stronger dependency on the liner mass per-unit-length $\mhat$ as compared to the shell entropy parameter $\alpharef$.  In other words, the shell entropy parameter $\alpharef$ seems to play a more prominent role in the determination of the plasma conditions at stagnation, but the mass per-unit-length becomes a more important parameter when considering integrated performance of a Z-pinch x-ray or neutron source.

\section{Discussion}
\label{sec:discussion}

\subsection{Comparison of the in-flight dynamics for cylindrical Z-pinch implosions and spherical laser-driven implosions}
\label{sec:sphere}

Let us now compare and contrast the in-flight dynamics of cylindrical Z-pinch implosions and spherical laser-driven implosions.  Regarding the latter, a common approximation is to consider the external pressure drive as constant in time.  Then, the momentum-conservation equation for a high--aspect-ratio spherical shell is
\begin{equation}
    m_{\rm shell} \frac{\dm^2 R}{\dm t^2}
        = - 4 \pi R^2 P_{\rm sb},
    \label{eq:sphere:R}
\end{equation}
where $m_{\rm shell}$ is the mass of the shell and $P_{\rm sb}$ is the pressure drive after shock breakout.  Integrating the equation above gives the energy equation:
\begin{equation}
    \left( \frac{\dm R}{\dm t} \right)^2 
        =   \frac{8 \pi P_{\rm sb}}{3 m_{\rm shell}} 
            \left( R_0^3 - R^3\right),
    \label{eq:sphere:E}
\end{equation}
which we shall use to calculate the shell Mach number.

During the isentropic-acceleration phase of a spherical implosion, the shell pressure, density, thickness, and IFAR obey\cite{basko2002}
\begin{equation}
	\begin{aligned}
	\frac{P}{\Psb} &= 1, 
	&
    \frac{D}{\Dsb} &= 1, \\
    \frac{\Delta}{\Deltasb}	&= \left(\frac{R}{R_0}\right)^{-2} , 
	&
	\frac{A}{\Asb}	&= \left(\frac{R}{R_0}\right)^{3}.
    \end{aligned}
    \label{eq:sphere:Q2}
\end{equation}
The in-flight Mach number is given by
\begin{equation}
	M^2 =  \frac{U^2}{\gamma \Psb/\Dsb}
        =  \frac{2A_{\rm sb}}{3(\gamma-1)} 
            \left( 1- \frac{A}{A_{\rm sb}} \right) ,
    \label{eq:sphere:M2}
\end{equation}
where we used $m_{\rm shell} \simeq 4 \pi \rho_0 R_0^2 \Delta_0$ and substituted \Eqs{eq:phase1:Dsb}--\eq{eq:phase1:Asb}, \eq{eq:sphere:E}, and \eq{eq:sphere:Q2}.

\begin{figure}
	\includegraphics[width=1\linewidth]{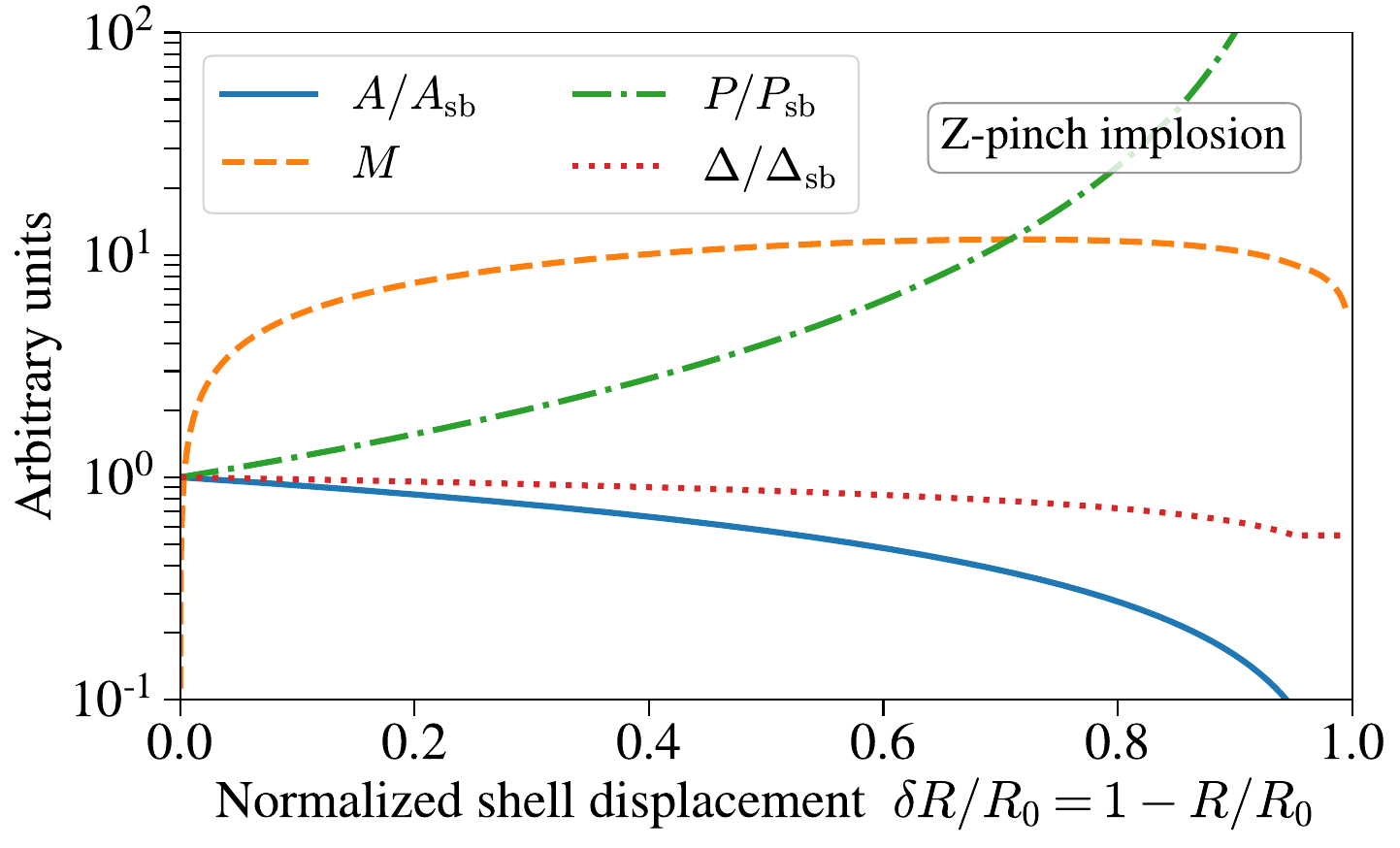}
	\includegraphics[width=1\linewidth]{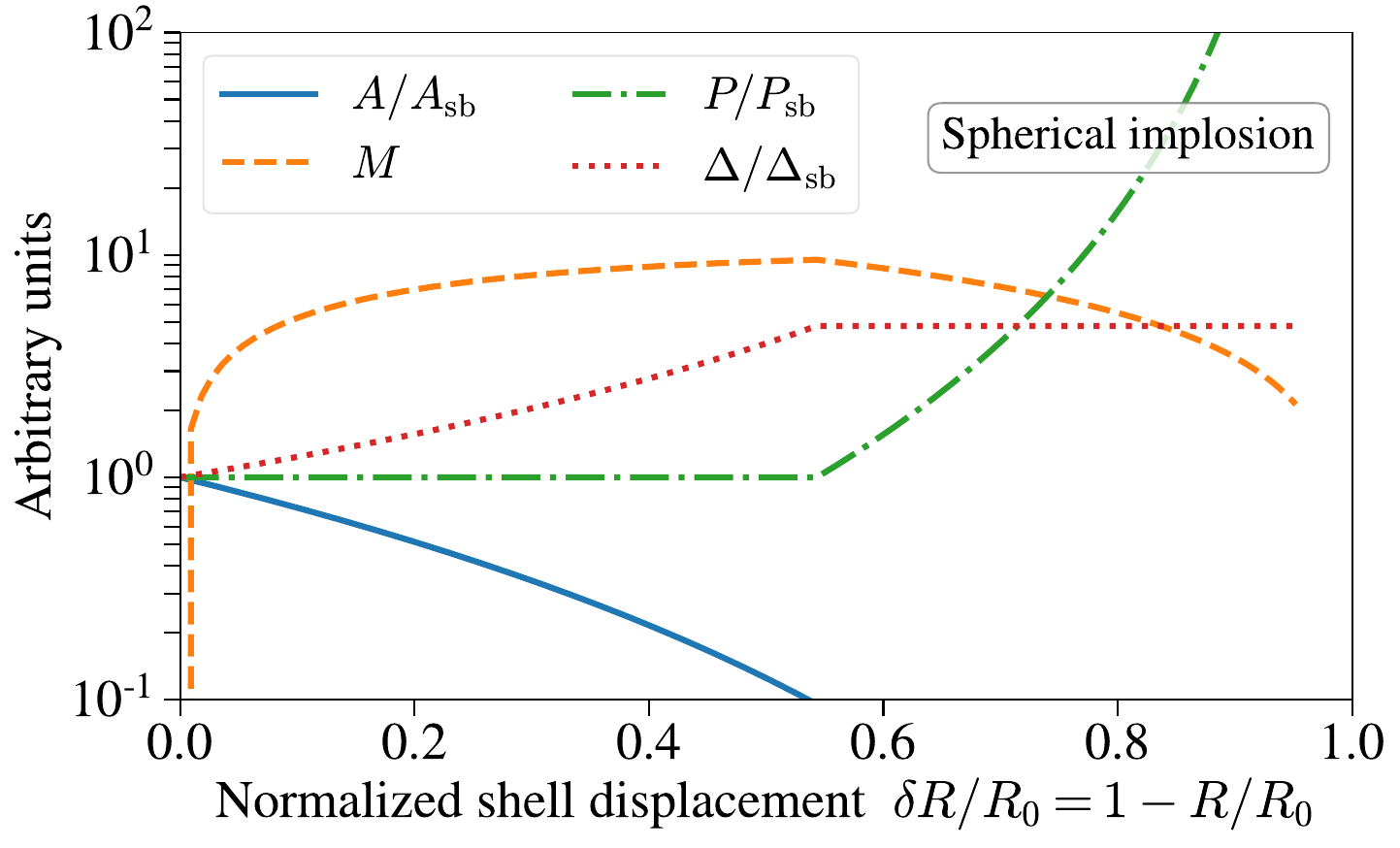}
	\caption{Evolution of the normalized aspect ratio $A/\Asb$, Mach number $M$, normalized pressure $P/\Psb$, and normalized shell thickness $\Delta/\Deltasb$ for a Z-pinch implosion (top) and a spherical implosion (bottom).  For both cases, we consider $\Asb=10^2$ and $\gamma=5/3$.}
	\label{fig:trajectories}
\end{figure}

Let us now compare \Eqs{eq:sphere:Q2} and \eq{eq:sphere:M2} for a spherical implosion to the results found in \Eqs{eq:phase2:Q} and \eq{eq:phase2:M2} for a Z-pinch implosion.  Due to the constant pressure drive and conservation of the shell adiabat, the shell pressure and density do not change in the spherical case.    In contrast, the liner density for a Z-pinch increases as the implosion proceeds.  This represents an advantage for Z pinches since this effect is beneficial for increasing the in-flight ram pressure of the shell.  The evolution of the shell pressure is illustrated in the green dot-dashed curves in \Fig{fig:trajectories} for both cases.  For this comparison, we set the initial aspect ratio at shock breakout to be equal for both cases.  It is worth mentioning that, for the spherical case, the shell thickness increases as $R^{-2}$, and the shell IFAR decreases as $R^3$.  These trends are advantageous for spherical implosions compared to Z-pinches since the expansion of the shell width and the subsequent faster decrease in IFAR are beneficial for mitigating the feedthrough of the Rayleigh--Taylor instability.  The evolution of the shell thickness and the IFAR are illustrated by the solid blue lines and red dotted lines in \Fig{fig:trajectories}, respectively.

After the isentropic-acceleration phase, an imploding spherical shell reaches the ``coasting" phase, where $T_{\rm ex}>T_{\rm imp}$ and the shell thickness remains approximately constant.  Here the pressure, density, thickness, IFAR, and Mach number obey\cite{basko2002}
\begin{equation}
	\begin{aligned}
	\frac{P}{P_\star} 
        &= \left(\frac{R}{\Rstar}\right)^{-2\gamma},
	&
	\frac{D}{D_\star} 
        &= \left(\frac{R}{\Rstar}\right)^{-2}, 
	&
	\Delta	= \Delta_\star , 
	\\
	\frac{A}{\Astar}	
    &    = \left(\frac{R}{\Rstar}\right), 
	&
	 \frac{M}{\Mstar} 
        &\simeq \left(\frac{R}{\Rstar}\right)^{\gamma-1}. 
    \end{aligned}
    \label{eq:sphere:Q3}
\end{equation}
During this stage, the shell density amplifies as $R^{-2}$ due to spherical-convergence effects.  This leads to large amplification of the in-flight ram pressure and high pressures at stagnation. The sudden jump in the shell pressure is observed in \Fig{fig:trajectories}~{\color{blue}(bottom)}.

An important difference shown in \Fig{fig:trajectories} is that, for spherical implosions, the implosion-at-constant-thickness phase occurs relatively early when $\Rstar/R_0 \simeq 0.45$ for the example given in \Fig{fig:trajectories}.  For Z-pinch implosions, this stage begins when $\Rstar/R_0 \simeq 0.05$, meaning that the the transition $T_{\rm ex}>T_{\rm imp}$ occurs right before the moment of stagnation (see \Fig{fig:Rstar}).  This noticeable difference between the two imploding systems can be explained by the magnetic-drive of Z pinches, where the ever increasing magnetic pressure increases the characteristic sound-propagation speed inside the shell and delays the transition to the constant-thickness phase.

\begin{figure}
	\includegraphics[width=1\linewidth]{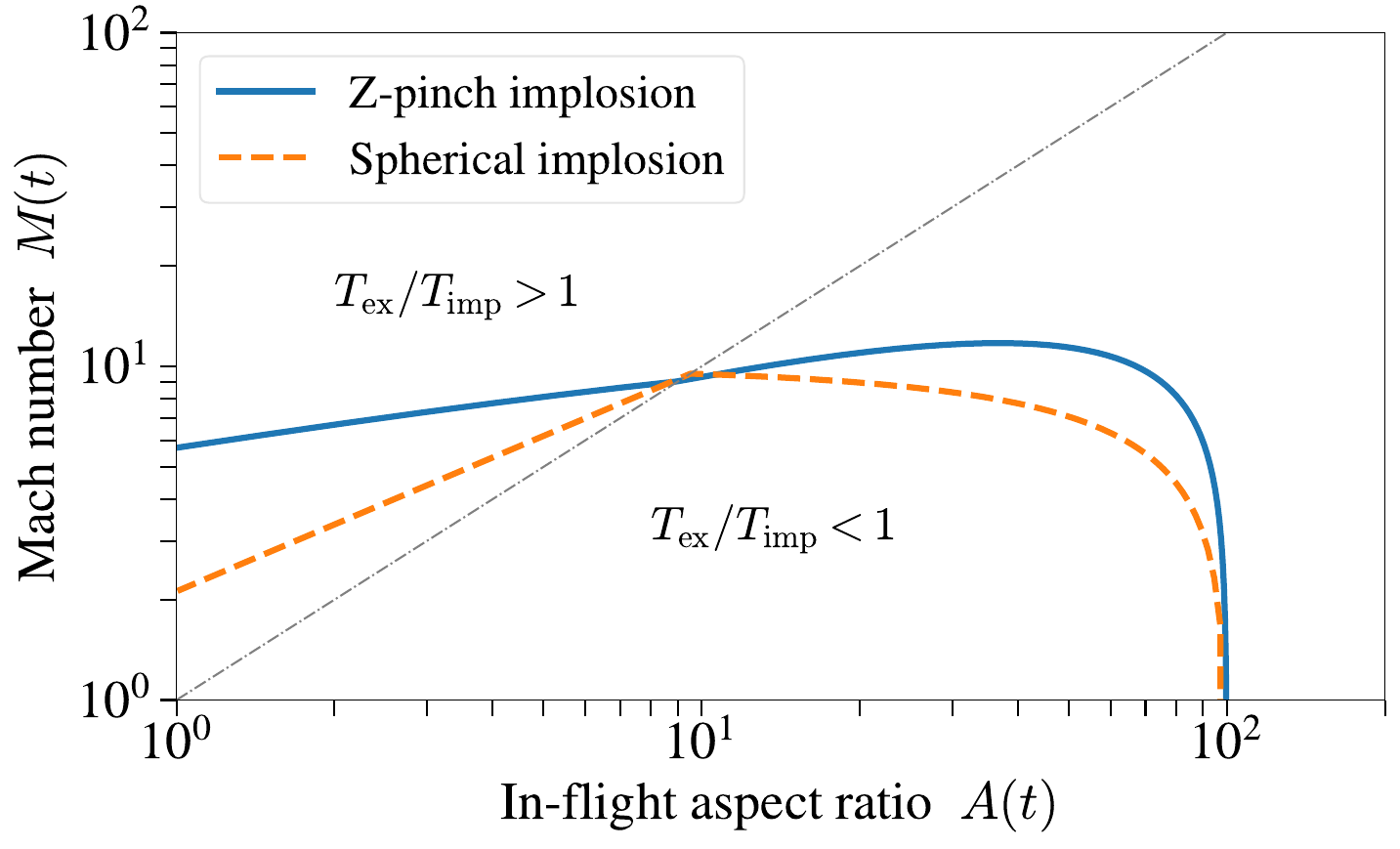}
	\caption{Implosion trajectories in the $(A,M)$ space for a magnetically-driven cylindrical Z pinch and a laser-driven spherical implosion. For both cases, we consider $\Asb=10^2$ and $\gamma=5/3$.}
	\label{fig:comparison}
\end{figure}

To conclude this section, in \Fig{fig:comparison}, we compare the implosion trajectories in the $(A,M)$ parametric plane for a cylindrical Z-pinch and a spherical laser-driven shell. Regarding the Mach number $M$, the Z-pinch case reaches a maximum during the acceleration phase but subsequently decreases as the IFAR decreases. Conversely, the Mach number for spherical implosions increases as the IFAR decreases when $T_{\rm ex}/T_{\rm im}<1$.  In \Fig{fig:comparison}, the transition $T_{\rm ex}/T_{\rm im}=1$ occurs at $A_\star \simeq 9$ for both cases.  However, it is important to note that this observation pertains specifically to the example utilizing $\Asb=100$ and $\gamma=5/3$ and should not be considered as a general property.  When $T_{\rm ex}/T_{\rm im}>1$, the spherical implosion case exhibits a more pronounced decrease in the Mach number. This phenomenon arises because, in spherical implosions, once the radius of the shell diminishes sufficiently, the acceleration of the shell becomes negligible. Consequently, the shell velocity remains relatively constant, while the shell pressure and density increase significantly during the coasting phase. In contrast, the Z-pinch continues to be accelerated due to the increasing magnetic pressure drive, which partially offsets the rising shell density and pressure during the latter stages of the implosion.  Despite the differences in the implosion trajectories in the $(A,M)$ plane for spherical and Z-pinch imploding systems, our estimates indicate that both systems converge to final Mach number values greater than unity at stagnation when starting with the same in-flight aspect ratios.  Therefore, we conclude that the stagnation pressure of a spherical or Z-pinch implosion is a result of an amplification of the in-flight hydrodynamic (or ram) pressure prior to void closure, not of the pressure source driving the implosion.

\subsection{Similarity scaling Z-pinch implosions with respect to peak current}
\label{sec:CS}

From the results presented in Secs.~\ref{sec:scaling} and \ref{sec:perf}, we may also determine the extrapolation scaling laws for the stagnation conditions and performance metrics when increasing the peak current driving a Z pinch.  One convenient scaling strategy presented in \Refs{Schmit:2020jd,ruiz2023,ruiz2023a,ruiz2023b} is to scale Z-pinch implosions so that dynamic similarity is maintained between the baseline and scaled load designs.  To achieve this, one strategy is to scale the target design parameters such that the implosion time $\Delta T$ remains constant.  Maintaining $\Delta T$ invariant in \Eq{eq:kinematics:DeltaT} when increasing the peak current leads to the scaling relation $\mhat R_0^2 \propto \I^2$.  In the high--aspect-ratio limit, the scaling law above becomes
\begin{equation}
	R_0 \propto \frac{\I^{1/2} A_0^{1/4}}{\rho_0^{1/4}}
		\propto \I^{1/2} A_0^{1/4}.
	\label{eq:CS:scaling}
\end{equation} 
In what follows, we shall consider that the liner material is left unchanged when scaling up in current.  Thus, the initial liner density $\rho_0$ is considered constant.  

\subsubsection{Current scaling while holding constant the initial aspect ratio $\boldsymbol{A_0}$}
\label{sec:CS1}

When scaling with respect to peak current while keeping the implosion time constant, there are two possibilities worth considering.  The first option involves increasing the liner radius while maintaining the initial liner aspect ratio $A_0$ constant.  Such scaling strategy may be more appropriate for wire-array implosions, whose effective IFARs are higher compared to solid metallic liner implosions, such as MagLIF.\cite{Slutz:2010hd,Gomez:2014eta,Gomez:2019bg,YagerElorriaga:2022cp}  Upon using \Eq{eq:CS:scaling}, we find that the initial liner radius, the mass per-unit-length, and the characteristic implosion velocity scale as
\begin{equation}
	R_0 \propto \Uref \propto \I^{1/2}, \qquad
	\mhat \propto \I.
	\label{eq:CS:scalingA0}
\end{equation}

We insert \Eqs{eq:CS:scalingA0} into the asymptotic scaling laws found in Secs.~\ref{sec:scaling} and \ref{sec:perf} for the stagnation conditions and performance metrics.  The resulting extrapolation scaling laws are summarized in the third column of \Tab{tab:scaling}.  From the algebraic manipulations, the noteworthy results are the following:  (i)~the temperature at stagnation scales linearly with current $\smash{(\Tstag \propto \I)}$, (ii)~the stagnation pressure scales with current as $\smash{(\Pstag \propto \I^{5/2})}$, which is a much stronger scaling law compared to the $\smash{P_{\rm mag} \propto \I}$ scaling law for the magnetic-drive pressure, (iii)~the confinement time is expected to decrease even though the implosion time $\Delta T$ is conserved $\smash{(\tau_{\rm stag} \propto \I^{-3/4})}$, (iv)~the K-shell emission per-unit-length approximately scales as $\smash{(\widehat{\mc{P}}_{\rm bb} \propto \I^{2.95})}$, and (v)~the neutron yield per-unit-length scales strongly with current $\smash{(\widehat{Y} \propto \I^{5.75})}$.  Regarding the last point, it is interesting to note that the predicted scaling relation for the neutron yield is more optimistic than the typically quoted $\I^4$ scaling law in the Z-pinch literature.\cite{Velikovich:2007hq}

\begin{table*}
\caption{Asymptotic scaling laws for stagnation conditions, energetic quantities, and performance metrics of a high--aspect-ratio $(A\gg1)$ Z-pinch implosion.  A polytropic index of $\gamma=5/3$ is considered in the summarized results below.} 
\label{tab:scaling}
\begin{tabular}{ 
>{\centering\arraybackslash}m{0.3\linewidth} | 
>{\centering\arraybackslash} m{0.15\linewidth} |
>{\centering\arraybackslash} m{0.25\linewidth} |
>{\centering\arraybackslash} m{0.25\linewidth} 
}
\hline
\hline
Physical quantity  & Asymptotic scaling  
					& Current scaling at fixed initial aspect ratio $A_0$   
					& Current scaling at fixed in-flight aspect ratio $\Asb$ \\
\hline
\hline
Stagnation radius $\Rstag$
					& $\displaystyle\frac{\alpharef^{3/4} \mhat^{1/2}}{\Uref^{3/2}}$ 	
					& $\I^{-1/4}$
 					& $\I^{2/7}$ 				\\
\hline
Stagnation density $\Dstag$
					& $\displaystyle\frac{\Uref^3}{\alpharef^{3/2}}$ 	
					& $\I^{3/2}$
 					& $\I^{6/7}$ 				\\
\hline
Stagnation pressure $\Pstag$
					& $\displaystyle\frac{\Uref^5}{\alpharef^{3/2}}$ 	
					& $\I^{5/2}$
 					& $\I^{10/7}$				\\
\hline
Stagnation temperature $\Tstag$
					& $\displaystyle 
						{\color{white}\frac{A^2}{B^2}} 
						\Uref^2 {\color{white}\frac{A^2}{B^2}}$ 	
					& $\I $
 					& $\I^{4/7}$				\\
\hline
Areal density $\sigma_{\rm stag}$
					& $\displaystyle\frac{\mhat^{1/2} \Uref^{3/2}}{\alpharef^{3/4} }$ 	
					& $\I^{5/4}$
 					& $\I^{8/7}$			\\
\hline
Confinement time $\tau_{\rm stag} $
					& $\displaystyle \frac{\mhat^{1/2} \alpharef^{3/4} }{\Uref^{5/2}}$ 	
					& $\I^{-3/4}$
 					& \const				\\
\hline
\hline
Internal energy per-unit-length $\widehat{\mc{E}}_{\rm int}$
					& $\displaystyle{\color{white}\frac{A^2}{B^2}} 
						\mhat \Uref^2 {\color{white}\frac{A^2}{B^2}}$ 	
					& $\I^2 $
 					& $\I^2 $ 				\\
\hline
Kinetic energy per-unit-length $\widehat{\mc{E}}_{\rm kin}$
					& $\displaystyle{\color{white}\frac{A^2}{B^2}} 
						\mhat \Uref^2 {\color{white}\frac{A^2}{B^2}}$ 	
					& $\I^2 $
 					& $\I^2 $ 				\\
\hline
\hline
Bremsstrahlung emission per-unit-length $\widehat{\mc{P}}_{\rm ff}$
					& $\displaystyle \frac{\mhat^{3/2} \Uref^{3/2} }{\alpharef^{3/4}}$
					&  $\I^{2.25}$ 
					& $\I^{2.57}$ \\
\hline
X-ray line emission per-unit-length $\widehat{\mc{P}}_{\rm bb}$ (at $E_{mn}/k_{\rm B}T \simeq 1.7$)
					& $\displaystyle \frac{\mhat^{3/2} \Uref^{2.9} }{\alpharef^{3/4}}$
					&  $\I^{2.95}$ 
					& $\I^{2.97}$ \\
\hline
Neutron production per-unit-length $\widehat{Y}$ (at $3~{\rm keV} \leq T \leq 5~{\rm keV}$)
					& $\displaystyle \frac{\mhat^{3/2} \Uref^{17/2} }{\alpharef^{3/4}}$
					&  $\I^{5.75}$ 
					& $\I^{4.57}$ \\
\hline
\hline
\end{tabular}
\end{table*}

\subsubsection{Current scaling while holding constant the aspect ratio $\boldsymbol{\Asb}$ at shock breakout}
\label{sec:CS2}

The scaling strategy proposed in \Sec{sec:CS1} keeps the initial aspect ratio of the liner constant, which then leads to an aggressive scaling of the liner initial radius.  A more conservative approach designed to mitigate the effects of Rayleigh--Taylor instabilities is to scale the liner radius and mass such that the in-flight aspect ratio at shock breakout is maintained; \ie $\Asb=\const$  This scaling approach was proposed in \Refs{Schmit:2020jd,ruiz2023}.  From \Eqs{eq:phase1:Qsb} and \eq{eq:CS:scaling}, we find that the liner outer radius and mass per-unit-length obey
\begin{equation}
	\frac{R_0^4}{A_0} \propto \I^2, \qquad R_0^{2/\gamma} \propto \frac{1}{A_0^{1-1/\gamma}}.
\end{equation}
We consider a polytropic index of $\gamma=5/3$ and obtain
\begin{equation}
	R_0 \propto \Uref \propto \I^{2/7},
	\qquad
	\mhat \propto \I^{10/7}. 
	\label{eq:CS:scalingAsb}
\end{equation}
Comparing the scaling prescriptions in \Eqs{eq:CS:scalingAsb} to those in \Eqs{eq:CS:scalingA0}, we find that the scaled liners grow more slowly in radius and their mass increases at a faster rate.  In fact, with this scaling strategy, the liner initial aspect ratio decreases almost linearly with current: $\smash{A_0 \propto \I^{-6/7}}$.

We insert the scaling prescriptions in \Eqs{eq:CS:scalingAsb} into the asymptotic scaling laws found in Secs.~\ref{sec:scaling} and \ref{sec:perf} for the stagnation conditions and performance metrics.  The resulting extrapolation scaling laws are summarized in the fourth column of \Tab{tab:scaling}.  Comparing the results between the third and the fourth columns, we find that the extrapolation scaling laws using a fixed initial aspect ratio tend to be more favorable.  This occurs because the liner velocity increases more strongly in \Eqs{eq:CS:scalingA0}.  Nevertheless, the scaling law for the neutron yield following the second scaling strategy (with fixed in-flight aspect ratio) is still more optimistic than the typical $\I^4$ scaling law.  It is surprising that the second scaling strategy shows a more favorable scaling law for the generation of x-rays compared to the first strategy.  This is due to the strong dependency of x-ray emission on the mass per-unit-length, which increases more strongly in \Eqs{eq:CS:scalingAsb}.

An important observation to conclude this section is that, for both scaling strategies, the x-ray emission demonstrates a weaker dependence on peak current compared to the neutron yield. This suggests that, when  enhancing the peak current and employing similarity scaling of Z-pinch configurations, increasing the x-ray emission is harder than increasing the neutron yield.

\section{Conclusions}
\label{sec:conclusions}

In this work, we theoretically investigated the in-flight dynamics of a magnetically-driven, imploding cylindrical shell that stagnates onto itself upon collision on axis.  The converging flow of the Z-pinch is analyzed by considering the implosion trajectory in the $(A, M)$ parametric plane, where $A$ is the in-flight aspect ratio and $M$ is the implosion Mach number.  For an ideal implosion in the absence of instabilities, we derived the asymptotic scaling laws for hydrodynamic quantities (\eg density, temperature, and pressure) evaluated at stagnation as functions of target-design parameters.  We obtained the asymptotic scaling laws for various metrics measuring the performance of popular Z-pinch applications, including x-ray emission and neutron yield.

Our study suggests that the Z-pinch implosion velocity is the most important lever for enhancing the HED conditions of the Z-pinch plasma at stagnation.  For the hydrodynamic conditions at stagnation of the Z pinch plasma (\eg density, temperature, and pressure), the next most important parameter is the shell entropy parameter $\alpharef$.  When lowered, it can increase the density, pressure, and areal density of the Z pinch at stagnation.  For integrated performance metrics, such as x-ray emission and neutron yield, the mass per-unit-length $\mhat$ plays a more prominent role compared to the entropy parameter.

This work also compares the kinematics of cylindrical Z-pinch implosions and spherical laser-driven implosions. These systems differ in two major aspects. For Z pinches, the magnetic pressure driving the implosion increases as the shell converges on axis, leading to a \textit{continuously} increasing shell density and pressure in flight. Our analysis suggests that the implosion phase, during which the shell thickness remains constant, occurs relatively late in Z-pinch implosions (\ie at high convergence ratios).  Therefore, the magnetic-drive pressure in Z-pinch implosions plays an important role in establishing the final implosion velocity of the shell and the in-flight density profile.  Both factors contribute to the hydrodynamic (or ram) pressure and to the final stagnation pressure.

For spherical implosions, the drive pressure is approximately constant, and the shell acceleration weakens as the surface area of the spherical shell decreases.  Thus, it becomes more difficult to accelerate a spherical shell once the convergence ratio is greater than 2 or 3.  Since the drive pressure is approximately constant, there are no gains in the hydrodynamic pressure due to increases in the shell density during the first phase of the implosion.  However, the coasting phase, where the shell thickness remains approximately constant, occurs relatively early in spherical implosions.  This phase is crucial for spherical implosions because the shell density and pressure amplify dramatically during this phase, increasing the hydrodynamic pressure and leading to high pressure at stagnation.  Despite the differences in the implosion trajectories in the $(A,M)$ parametric plane for spherical and Z-pinch imploding systems, both systems converge to similar final Mach number values at stagnation when starting with the same in-flight aspect ratios.

As a final remark, we emphasize that the scaling laws presented in this study are derived for an idealized one-dimensional Z-pinch implosion. A number of non-ideal effects are not considered in this analysis; for example, hydrodynamic instabilities, complex 2D and 3D hydrodynamic flows, finite-conductivity phenomena, and energy-loss mechanisms occurring in-flight during the implosion and at stagnation.  Incorporating these effects would undoubtedly alter the scaling exponents obtained in Secs.~\ref{sec:scaling} and \ref{sec:perf} for the stagnation conditions and performance metrics of a Z-pinch implosion.

For future research, we propose the following plan to test the scaling laws presented in this study against simulation results and experimental data. As an initial step, it would be beneficial to validate the scaling laws through 1D radiation-magneto-hydrodynamic simulations that incorporate finite-conductivity phenomena and energy-transport mechanisms. Subsequent efforts can focus on investigating the effects of hydrodynamic instabilities and complex multi-dimensional hydrodynamic flows, including the influence of residual kinetic energy on performance degradation.\cite{hurricane2020,foot:ruiz2024a} Finally, existing experimental data can be analyzed to uncover underlying scaling laws that reflect the parametric dependencies discussed in this work.

The data that support the findings of this study are available from the corresponding author upon reasonable request. 

Sandia National Laboratories is a multimission laboratory managed and operated by National Technology $\&$ Engineering Solutions of Sandia, LLC, a wholly owned subsidiary of Honeywell International Inc., for the U.S. Department of Energy's National Nuclear Security Administration under contract DE-NA0003525.  This paper describes objective technical results and analysis. Any subjective views or opinions that might be expressed in the paper do not necessarily represent the views of the U.S. Department of Energy or the U.S. Government.

\appendix

\section{Constant shell thickness during Phase 3 of the implosion}
\label{app:Delta}

The radial momentum equation with the magnetic pressure term included reads as
\begin{equation}
	\rho \left( \frac{\pd v_r}{\pd t} + v_r \frac{\pd}{\pd r} v_r \right) = - \frac{\pd p}{\pd r} - \frac{\pd p_m}{\pd r}.
\end{equation}
We write the radial velocity as $v_r(t,r) = \dot{R}(t)+\delta v_r (t,r)$.  Upon noting that $\rho \sim \mhat/(2\pi R\Delta)$ and $\pd_r p_m\sim P_m/\Delta$ and using \Eq{eq:kinematics:R}, we have
\begin{equation}
	\frac{\pd}{\pd t} \delta v_r + (\dot{R} + \delta v_r) \frac{\pd}{\pd r} \delta v_r 
		= - \frac{1}{\rho}\frac{\pd}{\pd r} p.
\end{equation}
The correction velocity scales as $\delta v_r \sim \dot{\Delta}$.  The temporal and radial derivatives scale with the implosion time and the shell thickness; that is, $\pd_t \sim T_{\rm imp}^{-1} = R/\dot{R}$ and $\pd_r \sim \Delta^{-1}$.  When noting that $p/\rho \sim P/D \sim C_s^2$, we have
\begin{equation}
	\frac{\dot{\Delta}}{R/\dot{R}} 
		+ \frac{\dot{R}\dot{\Delta}}{\Delta}
		+ \frac{\dot{\Delta}^2}{\Delta} 
		\sim \frac{C_s^2}{\Delta}.
\end{equation}
The second term dominates the third one since $\dot{\Delta}\ll \dot{R}$ for a high--aspect-ratio shell.  Therefore, we obtain the following balance relation:
\begin{equation}
	\underbrace{\frac{\dot{\Delta}}{R/\dot{R}} }_1
		+ \underbrace{\frac{\dot{R}\dot{\Delta}}{\Delta}}_2
		\sim \underbrace{\frac{C_s^2}{\Delta}}_3.
\end{equation}
If terms $\#1$ and $\#2$ dominate over $\#3$, then $\dot{\Delta}~\sim 0$; that is, the shell thickness remains constant.  If term $\#1$ is balanced by $\#3$, we have $\dot{\Delta} \sim (C_s^2/\Delta)(R/\dot{R})$.  When denoting the change in the liner thickness by $\delta \Delta \sim \dot{\Delta} \, T_{\rm imp} = \dot{\Delta} (R/\dot{R})$, we obtain
\begin{equation}
	\frac{\delta \Delta}{\Delta} 
		\sim \frac{C_s^2}{\dot{R}^2} \frac{R^2}{\Delta^2}
		= \frac{A^2}{M^2} \ll 1,
\end{equation}
where $A\ll M$ since the expansion time is much longer than the implosion time during Phase 3 of the implosion [see \Eq{eq:phase2:ratioT}].  Finally, if term $\#2$ is balanced by term $\#3$, we obtain $\dot{\Delta} \sim C_s^2/\dot{R}$.  Then, 
\begin{equation}
	\frac{\delta \Delta}{\Delta} 
		\sim \frac{C_s^2}{\dot{R}^2} \frac{R}{\Delta}
		= \frac{A}{M^2}
		< \frac{A}{M}
		\ll 1,
\end{equation}
where we used $M>1$.  Therefore, $\delta \Delta / \Delta \ll 1$; \ie the shell thickness remains approximately constant during Phase 3 of the implosion.


\begin{thebibliography}{61}
\bibitem[{foo({\natexlab{c}})}]{foot:simeq}
\bibinfo{note}{The ``asymptotically equal" symbol was used in
  \Eqs{eq:phase1:Psb} and \eq{eq:phase1:Asb} since the liner outer surface
  technically moves inwards during the time period when the shock traverses the
  liner. Here, we do not take into account this small displancement. Hence,
  $R_{\rm sb} \simeq R_0$.}
\end{thebibliography}

\end{document}